\documentclass[aps]{revtex4}
\def\pochh #1#2{{(#1)\raise-4pt\hbox{$\scriptstyle#2$}}}
\def\binom#1#2{\left(\begin{array}{c}#1\\#2\end{array}\right)}
\font\mas=msbm10 \font\mass=msbm7 \textfont8=\mas \scriptfont8=\mass
\def\masy{\fam8}
\def\ms#1{{\masy#1}}
\usepackage{graphicx} 
\usepackage{dcolumn} 

\begin {document}

\title {The possibility of the non-perturbative
 an-harmonic correction to Mehler's
 formula for propagator of the harmonic oscillator. }
\author{J. Boh\' a\v cik}
\email{bohacik@savba.sk} \affiliation{Institute of Physics, Slovak
Academy of Sciences, D\' ubravsk\' a cesta 9, 845 11 Bratislava,
Slovakia.}
\author{P. Pre\v snajder}
\email{presnajder@fmph.uniba.sk} \affiliation{Department of
Theoretical Physics and Physics Education, Faculty of Mathematics,
Physics and Informatics, Comenius University, Mlynsk\' a dolina F2,
842 48 Bratislava, Slovakia.}
\author{P. August\' in}
\email{peto1506@gmail.com} \affiliation{Department of Theoretical
Physics and Physics Education, Faculty of Mathematics, Physics and
Informatics, Comenius University, Mlynsk\' a dolina F2, 842 48
Bratislava, Slovakia.}

\begin{abstract}
We find the possibility of the non-perturbative
 an-harmonic correction to Mehler's
 formula for propagator of the harmonic oscillator.
We  evaluate the conditional
 Wiener measure functional integral with a term of the fourth order in
the exponent by an alternative method as in the conventional
perturbative approach. In contrast to the conventional perturbation
theory, we expand into  power series the term linear in the
integration variable in the exponent. We discuss the case, when the
starting point of the propagator is zero. We present the results in analytical form
 for positive and negative frequency.

 (PACS: 03.65.-w, 03.65Db, 03.65.d, 05.40.Jc)
\end{abstract}
\maketitle

\section{Introduction}

In the case of the Wiener path integral there are essentially two approaches for giving a strict definition:

- to define path integral via finite dimensional approximation. Then the path integral is appropriate continuum limit, when the number of time slices is going to infinity.

- to define the Wiener measure in the frame of axiomatic probabilistic measure theory as a Gaussian type measure on the set of trajectories.

Quantum theory is rather the language of integration
theory (not that probability theory) of the conditional Wiener measure theory.
We would like to use the path integral formalism to non-perturbative analytical description of an-harmonic
oscillator in quantum mechanics, eventually quantum field theory systems.
In quantum theory with imaginary time, we see the formal connection with path integral formalism
for Brownian motion. The main difference between classical physics description of the Brownian
motion as a random process
and the quantum physics description of particle motion via path integral
inhere in the interpretation of results. In classical physics
we interpret the results of the path integral as a probability of displacement of the particle from position $i$
to position $f.$ In quantum theory we evaluate the amplitude of a propagation of the particle by path integral,
and this should not be confused with statistical probability of underlying Brownian motion.

The transition probability for Brownian particle under external harmonic oscillator force
is proportional to Mehler's formula:
\begin {equation}
W(x_i, t_i; x_f, t_f)= \left(\frac{k}{2\pi \sinh(\nu)}\right)^{1/2}
\exp\left\{ -\frac{k(x_i^2 + x_f^2)}{2\tanh(\nu)} + \frac{k x_i x_f}{\sinh(\nu)}\right\}.
\label{mehler}
\end {equation}

This formula
was derived by F.G. Mehler (1866) who investigated the diffusion equation in the presence of
harmonic oscillator force, i.e., with unit mass harmonic oscillator Hamiltonian
\begin {equation}
H = -\frac{1}{2} \triangle + \frac{1}{2} k^2 x^2,\ \nu = k(\tau_f - \tau_i).
\end {equation}
on the right side of diffusion equation\cite{mehler}.
The same result was derived for the probability of a stochastic movement of Brownian particle in external harmonic potential using conditional measure Wiener path integral methods, see e.g., Hille\cite{hille}, Doob\cite{dob}.
Alternatively, equation (\ref{mehler}) can be obtained also as
propagator of harmonic oscillator in quantum mechanics (i.e. with $\tau$ replaced by $i t$). Feynman obtained it in
1948 within his path integral approach to quantum mechanics, see Feynman \cite{fey}, Feynman - Hibbs \cite{feynm}.

For harmonic potential the transition amplitude in quantum mechanics is defined by the formula:
$$W_{QM}(q_i, t_i; q_f, t_f) = <q_f|\exp{\{-\frac{i}{\hbar}(t_f - t_i)H_0\}}|q_i>, H_0 = \frac{1}{2m}\hat{p}^2 + \frac{1}{2}k^2q^2.$$
For imaginary time $t \rightarrow -i\tau$ the corresponding formula reads:
$$W_E(q_i, \tau_i; q_f, \tau_f) = <q_f|\exp{\{-\frac{1}{\hbar}(\tau_f - \tau_i)H_0\}}|q_i>,\ H_0 = \frac{1}{2m}\hat{p}^2 + \frac{1}{2}k^2q^2.$$

For a general potential $V(x)$ Feynman expressed the quantum mechanical transition amplitude as the path integral of the following form:
\begin {equation}
W_{QM}(q_i, t_i; q_f, t_f) =  \int_{all\ paths}\ \prod_{\tau=t_i}^{t_f}\frac{dx(t)}{\sqrt{\frac{2i\pi\hbar}{m}dt}}
\exp{\left\{\frac{i}{\hbar}\int_{t_i}^{t_f}\ dt\ \left(\frac{1}{2}m\dot{x}^2 - V(x)\right)\right\}}
\label{fkr}
\end {equation}
Later Kac  rigorously justified the imaginary time analog of Feynman path integral, see\cite{kac, jaffe}:
\begin {eqnarray}
& &W_{E}(q_i, \tau_i; q_f, \tau_f) = <q_f|\exp{\{-(\tau_f - \tau_i)H\}}|q_i> = \nonumber \\
&=& \int_{all\ paths}\ \prod_{\tau=\tau_i}^{\tau_f}\frac{dx(\tau)}{\sqrt{\frac{2\pi\hbar}{m}d\tau}}
\exp{\left\{-\frac{1}{\hbar}\int_{\tau_i}^{\tau_f}\ d\tau\ \left(\frac{1}{2}m\dot{x}^2 + V(x)\right)\right\}}
\label{fki}
\end {eqnarray}
for a broad class of potentials $V(x).$
The transition amplitudes defined by path integrals (\ref{fkr}), (\ref{fki}) are known as Feynman-Kac formulas.
Let us remark that only $W_E$ rigorously correspond to the theory of Wiener path integrals. The results for
$W_{QM}$ is obtained by analytical continuation to imaginary time, the procedure is known as Wick rotation.
It must be stressed that formula (\ref{fki}) for $W_E$ can be interpreted as analytically continued propagator of the quantum particle in the potential $V(x)$  but also as the probability of the Brownian particle in the external field, i. e. generalized Markov random proces. Similarity of such descriptions enables to use the methods of the classical statistics physics.

The transition probability, or imaginary time propagator can be represented by conditional measure Wiener path integral, defined by the continuum limit of time-sliced finite dimensional integral:
$$W_E(q_i, \tau_i; q_f, \tau_f) = \lim_{N \rightarrow \infty}\ W_N(q_i, \tau_i; q_f, \tau_f),$$
where
$$W_N(q_i, \tau_i; q_f, \tau_f) = \left(\frac{m}{2\pi \ \epsilon}\right)^{\frac{N}{2}}\int\ \prod_{j=1}^{N-1}\ dq_j
\exp{\{-S_N(q_i, \tau_i; q_f, \tau_f)\}}$$
and
$$S_N(q_i, \tau_i; q_f, \tau_f) = \epsilon\ \sum_{j=1}^N\ \left(\frac{m}{2}\left(\frac{q_j - q_{j-1}}{\epsilon}\right)^2 +
V\left(\frac{q_j + q_{j+1}}{2}\right)\right), \ \epsilon = \frac{\tau_f - \tau_i}{N}.$$

The path integral approach is frequently used in quantum mechanics and quantum field theory as it allows effectively derive/incorporate standard perturbative expansions and even indicate steps beyond perturbative methods, see e.g.,
Roepstorff\cite{gert}, Das\cite{das}, Chaichian - Demichev\cite{dem}. However, there are very few path integrals that allow explicit evaluation. Such are e.g., systems of harmonic oscillators and free (relativistic or Euclidean fields - the corresponding transition probabilities/amplitudes represent multidimensional generalizations of Mehler's formula.

Our aim is to evaluate the transition probability
for motion of the Brownian particle in a quartic an-harmonic external potential
given by a conditional measure Wiener integral, in the quantum mechanics in imaginary time formalism such system correspond to the symmetric an-harmonic oscillator. There are various approximative or numerical estimates of various quantities, e.g., eigen-energies of the systems, that go beyond standard perturbative methods. However, to our best knowledge, there is little known directly about the an-harmonic oscillator transition probability (propagator).

\noindent
Bellow we shall present a non-perturbative evaluation of conditional measure Wiener path integral
with quartic addition to the harmonic oscillator potential. There is no reason to assume that this an-harmonicity is small.
Let us briefly to describe the idea of the the evaluation of
the finite dimensional integral which was in details explained in
our previous article\cite{my}.

Let us first consider the one dimensional integral with four order
term in the exponent which will frequently appears in the sequel:
\begin {equation}
J_1=\int\limits _{-\infty}^{+\infty}\;dx\;\exp\{-(a x^4+b x^2+c x)\}
 \label {int1d}
\end {equation}
where $Re\: a>0$.
This integral is not given by (simple) formula. However, $J_1=J_1(a,b,c)$ is an entire function
for any complex values
 of $b$ and $c$, since there exist all integrals
 \[ \partial_c^n\partial_b^m J_1(a,b,c)=(-1)^{n+m}
 \int\limits _{-\infty}^{+\infty}
 \;dx\;x^{2m+n}\exp\{-(a x^4+bx^2+cx)\}
 \]
Consequently, the power expansions of $J_1=J_1(a,b,c)$ in $c$ and/or
$b$ has an infinite radius of convergence (and in particular they
are uniformly convergent on any compact set of values of $c$ and/or
$b$). Let us now consider the power expansion in $c$ which we shall
 use:
\begin {equation}
J_1=\sum\limits _{n=0} ^{\infty} \frac{(-c)^n} {n!}\int\limits
_{-\infty}^{+\infty}\;dx\;x^n\exp\{-(a x^4+b x^2)\}
 \label{sim1}
\end {equation}
The integrals here appearing can be expressed in terms of the
parabolic cylinder function $D_{\nu}(z)$, $\nu=-m-1/2$, (see, for
instance, \cite{prud},\cite{bateman}). For $n$ odd, due to symmetry
of the integrand the integrals (\ref{sim1}) are zero, for $n$ even, $n=2m$ we
have:

 \begin {equation}
 J_1=\frac{\Gamma(1/2)}{(2a)^{1/4}}\sum\limits _{m=0} ^{\infty}
 \frac{(\xi)^m}{m!}e^{z^2/4}D_{-m-1/2}(z)
 \label{s2}\ ,\ \xi=\frac{c^2}{4\sqrt{2a}}\ ,\ z=\frac{b}{\sqrt{2a}}
\end {equation}
This sum is convergent for any values of $c$, $b$ and $a>0$.

We apply this procedure to evaluation of the $N$ dimensional
integral in Section II and in the Appendix A, where we find the
precise result expressed in the form of the parabolic cylinder
functions, but we have still $N-1$ fold summations in Eq. (\ref{afindim3})
as the consequence of the Taylor's expansions during the evaluation. To
treat these summations we adopt the only approximations in our
calculation. How the procedure of such summations was done is
described in Section II and Appendix B, the result is Eq.
(\ref{arecur104}).

In Section III and in the Appendices C, D, E, F we evaluate the
continuum limit of the $N$ dimensional integral. The final formula
for the conditional Wiener measure path integral with a term
of the fourth order in the exponent (see Eq. (\ref{arecur104})) is product of
the Mehler's formula for harmonic oscillator (\ref{ch304})  with fixed  start point
zero and the an-harmonic correction to this formula
(\ref{ch342}). Our result, contrary to the conventional perturbative approach,
describe the propagator for an-harmonic oscillator for the positive or negative
 frequency term (in our model the
parameter $b$). In Section IV we show the evaluation of the
non-perturbative correction to the exponential factor of Mehler's formula for the harmonic oscillator.

\section{Evaluation of the path integral}

We are going to evaluate the path integral in phase space formally
written as \cite{das}, \cite{dem}
$$\mathcal{W} = \int[\mathcal{D}\pi(\tau)] [\mathcal{D}\varphi(\tau)]\exp (-\mathcal{S}[\pi, \varphi])\ .$$
\noindent
where $\pi(\tau)$ and $\varphi(\tau)$ are the phase space coordinates and  $\mathcal{S}[\pi, \varphi]$ denotes the corresponding Euclidean action.
We suppose that the  quantity $\mathcal{S}[\pi, \varphi])$ is quadratic in the variable $\pi(\tau).$
Performing the Gaussian integration over this variable $\pi(\tau)$ we obtain the continuum \textit{conditional Wiener measure}
path integral defined as
\begin {equation}
\mathcal{W} = \int\ \left[D\varphi(\tau)\right]\ \exp{(-E[\varphi])},
\label{pi1}
\end {equation}
where
\begin {equation}
E[\varphi] = \int\limits_0^{\beta}\ d\tau\left[c/2\left(\frac{\partial
\varphi(\tau)}{\partial \tau}\right)^2
+b\varphi(\tau)^2+a\varphi(\tau)^4\right].
\end {equation}

\noindent
In the \textit{conditional Wiener measure} path integral, by definition the values $\varphi(0)=x_i$ and
$\varphi(\beta)=x_f$ are fixed.
The path integral (\ref{pi1}) can be defined as the limit of the finite dimensional integral   \cite{das}:

\begin {equation}
\mathcal{W}_{N}=
\left(\frac{1}{\sqrt{\frac{2\pi\triangle}{c}}}\right)^N
 \int\limits
_{-\infty}^{+\infty} \prod \limits _{i=1}^{N-1} d\varphi_i
\exp(-E_{N}) , \label{afindim}
\end {equation}
with
\begin {equation}
E_{N}= \sum\limits _{i=1}^N \triangle\left[c/2
\left(\frac{\varphi_i-\varphi_{i-1}}{\triangle}\right)^2
+b\varphi_i^2+a\varphi_i^4\right], \label{afindim1}
\end {equation}
representing the standard time-slice discretization of $E[\varphi].$
The factor
$$\left(\frac{2\pi\triangle}{c}\right)^{-N/2}$$
is the result of the gaussian integration over conjugate momentum
variable $\pi(\tau)$ and is the same for the \textit{conditional Wiener
measure} as well as \textit{unconditional Wiener measure} path
integrals \cite{das}.
The \textit{conditional Wiener measure} path integral is
defined by  limit
$$\mathcal{W} = \lim_{N\rightarrow\infty}\ \mathcal{W}_{N}.$$

To simplify evaluation  we fix the initial point $\varphi(0)=x_i=0.$
Performing successively all one dimensional integrals in (\ref{afindim}) we are all the time dealing with integrals
of the form (\ref{sim1}).
The evaluation of the $N-1$ dimensional integral (\ref{afindim}) is treated in the Appendix A. The result reads:

\begin {eqnarray}
\mathcal{W}_{N}& = &
\left(\frac{2\pi\triangle}{c}\right)^{-1/2}\left[\sqrt{2\pi(1+\frac{b\triangle^2}{c})}\right]^{-N+1}
\sum\limits_{k_1=0}^{\infty}\sum\limits_{k_2=0}^{\infty} \cdots
\sum\limits_{k_{N-1}=0}^{\infty}\nonumber \\
&\times&
\prod\limits_{i=1}^{N-2}\left\{\frac{(1+\frac{b\triangle^2}{c})^{-2k_i}}{(2k_i)!}
\Gamma(k_{i-1}+k_i+1/2)\mathcal{D}_{-k_{i-1}-k_i-1/2}(z)\right\}\label{afindim3}\\
&\times&
\frac{(1+\frac{b\triangle^2}{c})^{-k_{N-1}}}{(2k_{N-1})!}\left(\frac{c}{\triangle}
\varphi_N^2\right)^{k_{N-1}}\Gamma(k_{N-2}+k_{N-1}+1/2)\mathcal{D}_{-k_{N-2}-k_{N-1}-1/2}(z)\nonumber
\\
&\times&\exp{\left\{-a\triangle\varphi_N^4 -
(\frac{c}{2\triangle}+b\triangle)\varphi_N^2\right\}}. \nonumber
\end {eqnarray}
Here, $a, b, c, \varphi_N=x_f, \triangle=\beta/N$ are the constants of
the model, the variable $z$ is defined as
$$z = \frac{c\left(1+\frac{b\triangle^2}{c}\right)}{\sqrt{2a\triangle^3}},$$
and functions $\mathcal{D}_{-\nu-1/2}(z)$ are related to parabolic
cylinder functions $D_{-\nu-1/2}(z)$ by the relation:
$$\mathcal{D}_{-\nu-1/2}(z) = z^{\nu+1/2}\ \exp{\left\{\frac{z^2}{4}\right\}}D_{-\nu-1/2}(z).$$
The result in Eq. (\ref{afindim3}) is an exact expression, we did not use any  approximation
in evaluation. As we shown in  \cite{my} the multiple summations are uniformly convergent.

Our aim is to separate the multiple summations over $k_i'$s to the leading term and the remainder,
disappearing in the continuum limit $\triangle \rightarrow 0$ when $z\approx \triangle^{-3/2}.$
The individual summation over given $k_i$ in the product (\ref{afindim3}) is:
\begin {equation}
\sum\limits_{k_i=0}^{\infty}\left\{\frac{(1+\frac{b\triangle^2}{c})^{-2k_i}}{(2k_i)!}
\Gamma(k_{i-1}+k_i+1/2)\mathcal{D}_{-k_{i-1}-k_i-1/2}(z)
\Gamma(k_{i}+k_{i+1}+1/2)\mathcal{D}_{-k_{i}-k_{i+1}-1/2}(z)\right\}
\label{single}
\end {equation}
We shall divide this sum to leading part and the remainder:
$$\sum\limits_{j=0}^{\mathcal{J}}\; \frac{(-1)^j }{j!\;(2z^2)^j}
\sum_{k_i=0}^{\infty}\Gamma(k_{i-1}+k_i+1/2)\Gamma(k_{i}+k_{i+1}+2j+1/2)\mathcal{D}_{-k_{i-1}-k_i-1/2}(z) +
R(\mathcal{J}, K_0)$$

To do this task,  we introduce the first and the
last approximation in our calculation, when one of the parabolic
cylinder functions is replaced by asymptotic Poincar\' e - type
expansion \cite{temme}, \cite{temme2} of the parabolic cylinder
functions valid for finite index and large argument $z$:
\begin {equation}
\mathcal{D}_{-\nu-1/2}(z)\; \equiv \;
z^{\nu+1/2}\;e^{z^2/4}\;D_{-\nu-1/2}(z)\;=
\;\sum\limits_{j=0}^{\mathcal{J}}\; (-1)^j
\;\frac{\pochh{\nu+1/2}{2j}}{j!\;(2z^2)^j}
+\varepsilon_{\mathcal{J}}(\nu, z),
\label{poincare}
\end {equation}
where $\varepsilon_{\mathcal{J}}(\nu, z)$ is the remainder of Poincar\' e - type
expansion. The $\pochh{\nu}{k}=\nu(\nu+1)\cdots (\nu+k-1)$ is the Pochhammer symbol.
This asymptotic expansion is particularly useful in the continuum limit $\triangle \rightarrow 0$ when
$z\approx \triangle^{-3/2}$ and functions $\mathcal{D}_{-\nu-1/2}(z) \rightarrow 1.$
The first term in (\ref{poincare}) contributes to the leading part of (\ref{single}) whereas the second part
generates the remainder. The $k_i$ summations of the leading part
can be performed by the use of the Taylor expansion formula for parabolic cylinder functions \cite{bateman},
which takes the form
\begin {equation}
e^{x^2/4}\sum\limits_{k=0}^{\infty}\;
\frac{\pochh{\nu}{k}}{k!}\;t^k\;D_{-\nu-k}(x)\;=\;e^{(x-t)^2/4}\;D_{-\nu}\;(x-t),
\end {equation}
The estimate of the remainder for the leading part of (\ref{afindim3}) can be found in \cite{my}.
The detailed evaluation and discussion of the summation over indexes $k_i$
for the \textit{conditional Wiener measure} path integral  can be find in the Appendix B,
where was found for the leading term of $N-1$ dimensional integral (\ref{afindim3}) the expression:
\begin {eqnarray}
\mathcal{W}_{N}^{leading} &=&
\frac{1}{\sqrt{\left(\frac{2\pi\triangle}{c}\right) \prod
\limits_{i=0}^{N-2}2\omega_i(1+b\triangle^2/c)}}\exp{\left\{-a\triangle\varphi_N^4
- (\frac{c}{2\triangle}+b\triangle)\varphi_N^2\ +\ \xi\right\}}\nonumber \\
&\times& \sum\limits_{\nu=0}^{\mathcal{J}}\ (-1)^{\nu} \
\frac{1}{\nu!\;(2z^2)^{\nu}}\ \sum \limits_{p=0}^{2\nu}\ (\xi)^p\
\left(N-1\right)^{2\nu}_{p}
\label{arecur104}
\end {eqnarray}
The new symbols $\xi$ and $\omega_i$ are defined as:
$$\xi = \frac{1}{\omega_{N-2}}\
\frac{c}{4\triangle(1+\frac{b\triangle^2}{c})}\ \varphi_N^2,\ \omega_{i+1} =1- \frac{\sigma^2}{\omega_i}, \ \omega_0 = 1,$$
and expression $\left(N-1\right)^{2\nu}_{p}$ is defined by the recurrence relation in Appendix B.

\section{Path integral as the continuum limit of the $N-1$ dimensional integral.}

The evaluation of $\mathcal{W}_{N}$ was the target of the preceding
sections, where we have found for the leading term of
$N-1$ dimensional integral the relation (\ref{arecur104}).
In this section we will discuss its continuum limit. The continuum
limit of the first line in (\ref{arecur104}) is evaluated in the Appendices C and D,
where we found:

\begin {eqnarray}
& &\lim_{N\rightarrow\infty}
\frac{1}{\sqrt{\left(\frac{2\pi\triangle}{c}\right) \prod
\limits_{i=0}^{N-2}2\omega_i(1+b\triangle^2/c)}}\exp{\left\{-a\triangle\varphi_N^4
- (\frac{c}{2\triangle}+b\triangle)\varphi_N^2\ +\ \xi\right\}}\ =\ \nonumber \\
&=&\frac{1}{\sqrt{\frac{2\pi}{c}\ \frac{\sinh(\gamma \beta)}{\gamma}}}\
\exp{\left\{- \frac{c \gamma}{2}\coth{(\gamma \beta)}\varphi_N^2\right\}}\ ,\ \gamma = \sqrt{2b/c}\ .  \label{ch304}
\end {eqnarray}

Formula (\ref{ch304}) represents the Mehler's formula for
imaginary time \cite{das}, \cite{dem} for the propagator of the
harmonic oscillator with starting point zero and end point
$\varphi_N\ .$ The an-harmonic content of the oscillator is stored
in the continuum limit of second line in (\ref{arecur104}):

\begin {equation}
\sum\limits_{\nu=0}^{\mathcal{J}}\ (-1)^{\nu} \
\frac{1}{\nu!\;(2z^2)^{\nu}}\ \sum \limits_{p=0}^{2\nu}\ (\xi)^p\
\left(N-1\right)^{2\nu}_{p}\ , \label{ch305}
\end {equation}

\noindent
 where the variable $z$ was defined in the previous section as:
$$z = \frac{c\left(1+\frac{b\triangle^2}{c}\right)}{\sqrt{2a\triangle^3}}\ .$$
We see, that $z$ in the continuum limit $\triangle \rightarrow 0$ diverges as
$\triangle^{-3/2}\ .$ We rewrite Eq. (\ref{ch305})  in the form
\begin {equation}
\sum\limits_{\nu=0}^{\mathcal{J}}\ (-1)^{\nu} \
\frac{1}{\nu!\;(2z^2\triangle^3)^{\nu}}\ \left(\triangle^{3\nu}\sum \limits_{p=0}^{2\nu}\ (\xi)^p\
\left(N-1\right)^{2\nu}_{p}\right) \label{ch306}
\end {equation}

Now the term $(2z^2\triangle^3)^{\nu}$ is finite in the continuum limit and we are going to evaluate the continuum limit
of the expression:
\begin {equation}
 \triangle^{3\nu}\sum \limits_{p=0}^{2\nu}\ (\xi)^{2\nu-p}\
\left(N-1\right)^{2\nu}_{2\nu-p}\ . \label{ch307}
\end {equation}

The detailed evaluation of the above expression is done in the
Appendices F and G, here we summarize the final result only.

The an-harmonic correction to Mehler's formula defined in Eq. (\ref{ch306})
in the continuum limit reads:

\begin {equation}
\sum\limits_{\nu=0}^{\mathcal{J}}\ (-a)^{\nu}\
\sum \limits_{p=0}^{2\nu}\frac{c^{-p}}{\pochh{1/2}{(2\nu-p)}} \left(\frac{1}{ Q^2(\beta)}\varphi_N^2\right)^{2\nu-p}
\sum\limits_{\{m_1,\ldots,m_\nu\}\atop m_1+\ldots+m_\nu=p}
\prod_{j=1}^{\nu}\Sigma(m_j,j,p_j)I_{m_1,\cdots, m_{\nu}}(0)
\label{ch342}
\end {equation}
where
\begin {equation}
I_{m_1,\cdots, m_{\nu}}(\tau) = \int_{\tau}^{\beta}d\tau_1\ \int_{\tau_1}^{\beta}d\tau_2\cdots
\int_{\tau_{\nu-1}}^{\beta}d\tau_{\nu}\ \
d^{m_1}(\tau_1)d^{m_2}(\tau_2)\cdots\ d^{m_{\nu}}(\tau_{\nu})\ Q^4(\tau_{\nu})\  \cdots\ Q^4(\tau_2)\
Q^4(\tau_1)\ .\nonumber
\end {equation}
The multiple summations in (\ref{ch342})
means the sum over all groups of indices $0 \leq m_j \leq 4$ satisfying the condition $m_1+\ldots+m_\nu=p.$
The dependence of symbols $\Sigma(m_j,j,p_j)$ where $p_j=p-m_1-\cdots-m_j,$ on the values $m_j$ is given in the Table (\ref{Vseobecna sumacia}):
\begin{table}[h]
\begin{center}
    \begin{tabular}{|c|c|}\hline
    $m_{j}$ & $\Sigma(m_j,j,p_j)$\\ \hline
    0 & $(2(\nu-j)-p_{j}+1/2)(2(\nu-j)-p_{j}+3/2)$\\ \hline
    1 & $4(2(\nu-j)-p_{j}+1/2)(2(\nu-j)-p_{j}+3/4)$\\ \hline
    2 & $6(2(\nu-j)-p_{j})(2(\nu-j)-p_{j}-1)+9(2(\nu-j)-p_{j})+3/4$\\ \hline
    3 & $4(2(\nu-j)-p_{j}-1/4)(2(\nu-j)-p_{j})$\\ \hline
    4 & $(2(\nu-j)-p_{j}-1)(2(\nu-j)-p_{j})$\\ \hline
    \end{tabular}
    \caption{Values of $\Sigma(m_j,j,p_j)$ for  $m_{j}$}
    \label{Vseobecna sumacia}
    \end{center}
    \end{table}

Equation \ref{ch342} is the key formula. For any given $\mathcal{J}$ it gives the an-harmonic
correction as a finite sum. The integrals $I_{m_1,\cdots, m_{\nu}}(\tau)$ are analyzed in the next section,
where we derive various recurrence relations that allow us to analyze an-harmonic correctios
successively in the parameter $p$.
The symbols $d(\tau)$  and $Q(\tau),$ following the definitions in
Appendices, reads:
$$d(\tau) = \frac{1}{2\gamma}\left(\coth(\gamma \tau)-\coth(\gamma \beta)\right),\
Q(\tau) = 2\sinh(\gamma \tau),\ \gamma = \sqrt{\frac{2b}{c}}\ .$$

\section{The analysis of the an-harmonic correction.}

In this Section we shall show the evidence that the an-harmonic corrections in (\ref{ch342})
give a non-perturbative contribution to Mehler's formula for propagator of harmonic
oscillator. In order to extract as much information as possible
we interchange the order of finite summations in Eq. (\ref{ch342}):
$$\sum\limits_{\nu=0}^{\mathcal{J}}\ \sum \limits_{p=0}^{2\nu}\rightarrow \sum \limits_{p=0}^{2\mathcal{J}}\sum\limits_{\nu=\lfloor \frac{p+1}{2}\rfloor}^{\mathcal{J}}.$$
For finite value $p,$ and enough high $\nu$ the product
$$\prod_{i=1}^{\nu}\Sigma(m_i,i,p_i)$$
contains many terms with $m_i=0.$

Let $m_j\neq 0,$ $m_k\neq 0,$ and $m_i=0,$ for $j \leq i \leq k.$ As the result, $p_i=p_k.$
The product of this chain of $\Sigma(m_i,i,p_k)$ is:
$$\Pi(k,j,p_k)=\prod_{i=k+1}^{j-1}(2(\nu-i)-p_k+1/2)(2(\nu-i)-p_k+3/2)=\frac{\Gamma(2(\nu-k)-p_k+1/2)}{\Gamma(2(\nu-j)-p_k+5/2)}$$
Here we used the identity:
$$\Gamma(x)\Gamma(x+1/2)=\frac{\sqrt{\pi}}{2^{2x-1}}\Gamma(2x).$$
Let $m_{j_i}\neq 0$ for  $j_i = j_1,\cdots,j_{\mu}.$ Then
$$\prod_{i=1}^{\nu}\Sigma(m_i,i,p_i)=\Pi(0,j_1,p)\Sigma(m_{j_1},j_1,p_{j_1})\Pi(j_1,j_2,p_{j_1})\cdots\Sigma(m_{j_{\mu}},j_{\mu},0)\Pi(j_{\mu},\nu+1,0).$$
This expression can be rewritten to the form:
\begin {eqnarray}
& &\prod_{i=1}^{\nu}\Sigma(m_i,i,p_i)=\frac{\Gamma(2\nu -p + 1/2)}{\Gamma(1/2)}\
\frac{\Gamma(2(\nu-j_1)-p_{j_1}+1/2)}{\Gamma(2(\nu-j_1)-p+5/2)}\Sigma(m_{j_1},j_1,p_{j_1})\cdots\label{alg1}\\
& & \cdots \frac{\Gamma(2(\nu-j_i)-p_{j_i}+1/2)}{\Gamma(2(\nu-j_i)-p_{j_{i-1}}+5/2)}\Sigma(m_{j_i},j_i,p_{j_i})\cdots
\frac{\Gamma(2(\nu-j_{\mu})+1/2)}{\Gamma(2(\nu-j_{\mu})-p_{j_{\mu-1}}+5/2)}\Sigma(m_{j_{\mu}},j_{\mu},0)\nonumber
\end {eqnarray}
We can rewrite the expression (\ref{alg1}) as the product of the algebraic factors which depend on all $m_i\neq 0:$
\begin {equation}
 F(j_i,m_{j_i},p_{j_i}) = \frac{\Gamma(2(\nu-j_i)-p_{j_i}+1/2)}{\Gamma(2(\nu-j_i)-p_{j_{i}}- m_{j_i}+5/2)}\Sigma(m_{j_i},j_i,p_{j_i})
\end {equation}
In above definition the identity $p_{j_i} = p_{j_{i-1}} - m_{j_i}$ was used.

With this definition we can rewrite Eq. (\ref{alg1}) in the terms of nonzero $m_i$:
\begin {equation}
\prod_{i=1}^{\nu}\Sigma(m_i,i,p_i)= \pochh{1/2}{2\nu-p}\prod_{i=1}^{\mu}F(j_i,m_{j_i},p_{j_i})
\end {equation}
In the Table (\ref{Hodnoty algebra}) we summarize the dependence of the values of the algebraic factor $F(j_i,m_{j_i},p_{j_i})$  on values $m_{j_i}\neq 0.$

 \begin{table}[h]
    \begin{center}
    \begin{tabular}{|c|c|}\hline
    $m_{j_k}$ & $F(j_k, m_{j_k}, p_{j_k})$\\\hline
    1 & $4(2(\nu-j_k)-p_{j_{k}}+3/4)$\\ \hline
    2 & $6(2(\nu-j_k)-p_{j_{k}}+1/4+i/4)(2(\nu-j_k)-p_{j_{k}}+1/4-i/4)$\\\hline
    3 & $4(2(\nu-j_k)-p_{j_{k}})(2(\nu-j_k)-p_{j_{k}}-1/2)(2(\nu-j_k)-p_{j_{k}}-1/4)$\\\hline
    4 & $(2(\nu-j_k)-p_{j_{k}})(2(\nu-j_k)-p_{j_{k}}-1)(2(\nu-j_k)-p_{j_{k}}-1/2)(2(\nu-j_k)-p_{j_{k}}-3/2)$\\ \hline
    \end{tabular}
    \caption{The  dependence of the algebraic factor on values $m_{j_k}$}
    \label{Hodnoty algebra}
    \end{center}
    \end{table}

 We stress out the important and interesting characteristics of integrals at the form (\ref{ch342})
\begin {equation}
I_{m_1,\ldots,m_n}(\tau)= \int\limits_{\tau}^\beta d\tau_1 \int\limits_{\tau_1}^\beta d\tau_2 \ldots \int\limits_{\tau_{n-1}}^\beta d\tau_n J_{m_1}(\tau_1) \ldots J_{m_n}(\tau_n).
\end {equation}
Putting $\tau=0$ and $J_a(\tau) = d^a(\tau) Q^4(\tau)$ the connection to integrals in Eq. (\ref{ch342}) is evident.
The crucial feature is the identity:
\begin {equation}
I_{a,b}(\tau)+I_{b,a}(\tau) = I_a (\tau)I_b(\tau) \label{kluc}
\end {equation}
In proof in the second term we change the order of the integrations and then rename  the integration variables $x\leftrightarrow y:$
    \begin{eqnarray*}
    & &I_{a,b}(\tau)+I_{b,a}(\tau) = \int\limits_{\tau}^\beta dx \int\limits_x^\beta dy J_a(x)J_b(y) + \int\limits_{\tau}^\beta dx \int\limits_x^\beta dy J_b(x)J_a(y) = \\
    &=& \int\limits_{\tau}^\beta dx \int\limits_x^\beta dy J_a(x)J_b(y) + \int\limits_{\tau}^\beta dy \int\limits_y^\beta dx J_b(y)J_a(x) = \int\limits_{\tau}^\beta \int\limits_{\tau}^\beta dx dy J_a(x)J_b(y) = I_a(\tau) I_b(\tau)
    \end{eqnarray*}
    With identity (\ref{kluc}) we can prove various identities for the product of such integrals, as for instance:
    \begin{eqnarray*}
    & & I_{m_1,\ldots,m_{n-1}} I_m = \\
   &=& I_{m,m_1,\ldots,m_{n-1}}+I_{m_1,m,m_2,\ldots,m_{n-1}}+\ldots+I_{m_1,\ldots,m_{j-1},m,m_{j},\ldots,m_{n-1}}+\ldots +I_{m_1,\ldots,m_{n-1},m}
   \end{eqnarray*}
    As the result of the product of integral with one index $m$ and another integrals with $n-1$ indexes
     $m_1, m_2,\cdots, m_{n-1}$ we obtain $n$ terms with $n$ indexes each.
    The index $m$ runs over all positions in the group of $n$ indexes, the indexes $m_i$ don't permute among them.

    For purpose to evaluate the an-harmonic correction (\ref{ch342}) we consider product of integrals, where one of the integrals possesses
    $n$ indexes of the same value.
    By mathematical induction we obtain the well-known identity:
    \begin {equation}
    I\underbrace{_{a,\ldots,a}}_{n}(\tau)= \frac{I_a^{n}(\tau)}{(n)!}
    \end {equation}
    The application of the identity (\ref{kluc}) give:
    \begin {equation}
    I_{\alpha}(\tau) I\underbrace{_{a,\ldots,a}}_{n-1}(\tau)=\sum_{j=1}^n\ I\underbrace{_{a,\ldots,a,\alpha_j,a,\ldots,a}}_{n}(\tau)
    \label{p10}
    \end {equation}
    The subscript $j$ in index $\alpha_j$ indicate the position of index $\alpha$ among the indices of the integrals $I_{a,\ldots,a,\alpha_j,a,\ldots,a}\ .$
    For purposes to evaluate (\ref{ch342}) we need:
    \begin {equation}
    I_{\alpha,\beta}(\tau) I\underbrace{_{a,\ldots,a}}_{n-2}(\tau)=\sum_{j=1}^{n-1}\sum_{k=j+1}^n\ I\underbrace{_{a,\ldots,a,\alpha_j,a,\ldots,a,\beta_k,a,\ldots,a}}_{n}(\tau)
    \end {equation}
    We will meet the case, when $\beta=a$, then in above expression the integrals
    $I\underbrace{_{a,\ldots,a,\alpha_j,a,\ldots,a}}_{n}(\tau)$ are independent on the summation index $k$ and we obtain:
    \begin {equation}
    I_{\alpha,a}(\tau) I\underbrace{_{a,\ldots,a}}_{n-2}(\tau)=\sum_{j=1}^{n}(n-j) I\underbrace{_{a,\ldots,a,\alpha_j,a,\ldots,a}}_{n}(\tau)
    \label{p11}
    \end {equation}
    We expanded the summation over index $j$ up to $n$ by adding the zero term for $j=n.$
    The evaluation of the another useful relations is given to Appendix G.

    Applying this new notations for the an-harmonic correction (\ref{ch342}) we have:
\begin {eqnarray}
& &\sum\limits_{p=0}^{2\mathcal{J}}\ c^{-p} \sum\limits_{\nu=[\frac{p+1}{2}]}^{\mathcal{J}}(-a)^{\nu}\
 \left(X_N\right)^{2\nu-p}
\sum\limits_{\{m_{1},\ldots,m_{\nu}\}\atop m_{j_1}+\ldots+m_{j_{\mu}}=p}
\prod_{i=1}^{\mu}F(j_i,m_{j_i},p_{j_i})\label{ancor}  \\
&\times&\ I\underbrace{_{0,\ldots,0,m_{j_1},0,\ldots,0,m_{j_2},0,\ldots,0,m_{j_{\mu}},0,\cdots,0}}_{\nu}(0)\nonumber
\end {eqnarray}
For simplicity, the notation $$X_N = \frac{\varphi_N^2}{ Q^2(\beta)}$$ was in troduced.

   To analyze this result, we can see that for $p=0$ the contribution to (\ref{ch342}) can be red:
    \begin {equation}
    \sum\limits_{\nu=0}^{\mathcal{J}}(-a)^{\nu}\
 \left(X_N\right)^{2\nu}
    \frac{I_0^{\nu}(0)}{\nu!} \approx \exp\left\{-\frac{aI_0(0)\varphi_N^4}{Q^4(\beta)}\right\}\label{p0}
    \end {equation}
    For sufficiently large $\mathcal{J}$ the combination of the terms with $\nu > \mathcal{J}$
    $$\sum\limits_{\nu=\mathcal{J}+1}^{\infty}(-a)^{\nu}\
 \left(X_N\right)^{2\nu}\frac{I_0^{\nu}(0)}{\nu!} <
 \frac{\theta\frac{a\varphi^4_NI_0(0)}{Q^4(\beta)}}{\mathcal{J}\mathcal{J}!},\ 0<\theta<1 $$
 can be neglected with sufficient precision.

The contribution for $p=1$ can be red:
    \begin {equation}
    \sum\limits_{\nu=1}^{\mathcal{J}}(-a)^{\nu}\
    \left(X_N\right)^{2\nu-1}\sum_{j_1=1}^{\nu}F(j_1,1,0)
    I\underbrace{_{0,\ldots,0,1_{j_1},0,\ldots,0}}_{\nu}(0)
    \end {equation}
    Inserting $F(j_1,1,0)$ from Table \ref{Hodnoty algebra} we have:
    \begin {equation}
    \left(-a X_N\right)\sum\limits_{\nu=1}^{\mathcal{J}}(-a)^{\nu-1}\
    \left(X_N\right)^{2\nu-2}\sum_{j_1=1}^{\nu}
    (8(\nu-j_1)+3)I\underbrace{_{0,\ldots,0,1_{j_1},0,\ldots,0}}_{\nu}(0)
    \end {equation}
    Following to Eqs. (\ref{p10}),(\ref{p11}) we have:
     \begin {eqnarray}
    & &\left(-a X_N\right)\sum\limits_{\nu=1}^{\mathcal{J}}(-a)^{\nu-1}\
    \left(X_N\right)^{2\nu-2}
    \left(8I_{1,0}(0)\frac{I_0^{\nu-2}(0)}{(\nu-2)!}+3I_1(0)\frac{I_0^{\nu-1}(0)}{(\nu-1)!}\right)\approx\nonumber\\
    &\approx& \exp\left\{-\frac{aI_0(0)\varphi_N^4}{Q^4(\beta)}\right\}
    \left\{-3a\left(X_N\right)I_1(0)+8a^2 \left(X_N\right)^3 I_{1,0}(0)\right\}
    \label{r34}
    \end {eqnarray}

    For the contribution for $p=2$ we must take into account that that it is divided to two parts, one for the case when one $m_j=2$ and another when two $m_j=1,$ and $m_k=1$ are nonzero. We have:
    \begin {eqnarray}
    & &\sum\limits_{\nu=1}^{\mathcal{J}}(-a)^{\nu}\
    \left(X_N\right)^{2\nu-2}\left\{\sum_{j_1=1}^{\nu}F(j_1,2,0)
    I\underbrace{_{0,\ldots,0,2_{j_1},0,\ldots,0}}_{\nu}(0)\right. \nonumber\\
    &+&\left.\sum_{j_1=1}^{\nu-1}\sum_{j_2=j_1+11}^{\nu}F(j_1,1,1)F(j_2,1,0)
    I\underbrace{_{0,\ldots,0,1_{j_1},0,\ldots,0,1_{j_2},0,\ldots,0}}_{\nu}(0)\right\}
    \end {eqnarray}
    In the spirit of previous contributions we find for the contribution to the an-harmonicity correction for $p=2:$
    \begin{eqnarray}
    & &\exp\left\{-\frac{aI_0(0)\varphi_N^4}{Q^4(\beta)}\right\} \left\{3/4 (-a)I_2(0)
    + (-a)^2 \left(30 I_{2,0}(0) +21I_{1,1}(0)\right)\left(X_N\right)^{2}\right)+ \label{r36} \\
    &+&\left.(-a)^3\ (48I_{2,0,0}(0)+144I_{1,1,0}(0)+24I_{1,0,1}(0))\left(X_N\right)^{4}
    + 64 (-a)^4\ (I_{1,0,1,0}(0)+2I_{1,1,0,0}(0))\left(X_N\right)^{6}\right\}\nonumber
    \end{eqnarray}
    Calculations can be extended to any value of $p.$
    The common characteristics of all calculations is the universal non-perturbative exponential correction to Mehler's formula given by the exponential factor:
    $$\exp\left\{-\frac{aI_0(0)\varphi_N^4}{Q^4(\beta)}\right\}$$
    where:
    \begin {equation}
    \frac{I_0(0)}{Q^4(\beta)}=
    \frac{3\gamma\beta-4\cosh(\gamma\beta)\sinh(\gamma\beta)+\cosh^3(\gamma\beta)\sinh(\gamma\beta)+\cosh(\gamma\beta)\sinh^3(\gamma\beta)}
    {8\gamma\sinh^4(\gamma\beta)}\ .
    \end {equation}
    The second factor, given in the braces in Eq. \ref{r36}, is the $p$ dependent and for any $p$ is represented as polynomial of degree $2p$ in the variable $-a.$

\section{Conclusions.}

We presented an analytical method of evaluation of the conditional Wiener measure path integral with a four-th order term in the action.
Instead of the conventional perturbative evaluations we expand the linear part of kinetic term of the action. We obtain the analytical results representing the an-harmonic correction to the Mehler's formula for propagator of the harmonic oscillator. The most important result is the universal non-perturbative exponential correction to the exponent of the Mehler's formula. For the an-harmonic oscillator this exponencial factor can be red:
\begin {equation}
\exp{\left\{- \frac{c \gamma}{2}\coth{(\gamma \beta)}\varphi_N^2-a\frac{3\gamma\beta-4\cosh(\gamma\beta)\sinh(\gamma\beta)+\cosh^3(\gamma\beta)\sinh(\gamma\beta)+\cosh(\gamma\beta)\sinh^3(\gamma\beta)}
    {8\gamma\sinh^4(\gamma\beta)}\varphi_N^4\right\}}
    \label{con1}
\end {equation}
On the Fig.1 we can see the dependence of the exponential term of the an-harmonicity correction when  the parameter $b$ is positive, or negative.
The other an-harmonic corrections stand from polynomials of order $2p$ in variable $a$ that multiply the exponential factor in (\ref{r34}) and (\ref{r36}).
 We presented fully the corrections for $p=0,1,2$ but they can be found systematically for any $p$.
The interesting is the dependence of the Eq. (\ref{con1}) on the negative frequency (what can occur for $b<0.$). The exponential factor in Eq.(\ref{con1}) approaches $-\infty$ when $\gamma\beta=\beta\sqrt{2b/c} \rightarrow ik\pi.$ This means, that propagator for this particular $b, \beta$ vanishes and the particle is frozen in the origin $i$, because it cannot be able to propagate to any other point of the space.
\begin{figure}
  \includegraphics[width=9cm]{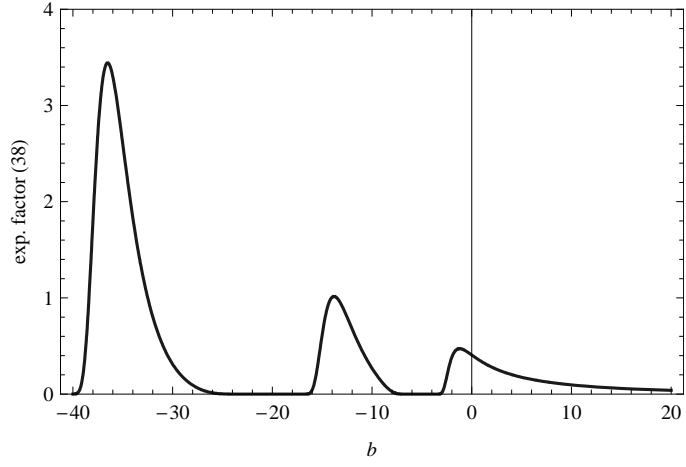}
 \caption{$b$ dependence of the exponential part of the Mehler's formula for an-harmonic oscillator (\ref{con1}) when the model parameters were fixed as $\beta = c = 1,\ a = 2.$
 }
\end{figure}

We can point to the evidence, that this correction play an important role in evaluations of the energy-levels of the an-harmonics oscillator. The energy levels of the harmonic oscillator correspond to the pole positions in the complex energy plane of the propagator obtained by Fourier transformation
$$\lim_{p\rightarrow 0}\int_{-\infty}^{\infty}dt\ e^{-iEt}\ \int_{-\infty}^{\infty}dx\ e^{-ipx}$$
of the propagator of the harmonic oscillator\cite{mehler} in the $x$ representation i. e. $W(x_i, 0; x_i, t).$
As the result of this Fourier transformation we have:
\begin {equation}
\frac{2 \pi}{\gamma}\tanh{\frac{E \pi}{\gamma}}
\end {equation}
The pole positions of this function appears for values

$$ Re E = 0,\ Im E = (n + 1/2)\gamma. $$

If we apply the same Fourier transformation to the propagator of the harmonic oscillator with fixed origin, we obtain the function with singularities fixed by formula:
$$\Gamma\left(1/4 + \frac{i E}{2 \gamma}\right)\Gamma\left(1/4 - \frac{i E}{2 \gamma}\right)$$

The pole positions of this function are:
$$ Re E = 0,\ Im E = \pm(2 n + 1/2)\gamma. $$
We see, that in this case we are capable to reproduce the even energy levels of the harmonic oscillator by the function corresponding to the propagator with fixed origin. Now, when we switch on the an-harmonics correction, we expect the shift of the pole position in the energy complex plane, that for small value of the an-harmonic parameter $a$ is represented by the first terms of expansion:
\begin {eqnarray}
& &\Gamma\left(1/4 + \frac{i E+\delta(a)}{2 \gamma}\right)\Gamma\left(1/4 - \frac{i E+\delta(a)}{2 \gamma}\right) = \\
&=&\Gamma\left(1/4 + \frac{i E}{2 \gamma}\right)\Gamma\left(1/4 - \frac{i E}{2 \gamma}\right)+
\Gamma\left(1/4 + \frac{i E}{2 \gamma}\right)\Gamma\left(1/4 - \frac{i E}{2 \gamma}\right)\left[\Psi\left(1/4 + \frac{i E}{2 \gamma}\right) -\Psi\left(1/4 - \frac{i E}{2 \gamma}\right)\right]\frac{\delta(a)}{2 \gamma} + \cdots \nonumber
\label{rozvoj}
\end {eqnarray}
When we approximate the an-harmonic propagator by simply adding the universal an-harmonic correction (39) to the propagator of the harmonic oscillator formula (1) with initial position $x_i=0$ (i.e. fixed at origin):
\begin {equation}
\frac{1}{\sqrt{\frac{2\pi}{c}\ \frac{\sinh(\gamma \beta)}{\gamma}}}\
\exp{\left\{- \frac{c \gamma}{2}\coth{(\gamma \beta)}\varphi_N^2 - \frac{a I_0(0)}{Q^4(\beta)}\varphi_N^4\right\}}
\label{prp}
\end {equation}
by the first part of Fourier transformation
$$\lim_{p\rightarrow 0}\ \int_{-\infty}^{\infty}dx\ e^{-ipx}$$
we have:
$$\frac{1}{\sqrt{\frac{2\pi}{c}\ \frac{\sinh(\gamma \beta)}{\gamma}}}\frac{\Gamma(1/2)}{\sqrt{(\frac{c\gamma}{2} \coth(\gamma \beta))}}\ \mathcal{D}_{-1/2}(z),\
z = \frac{\frac{c\gamma}{2} \coth(\gamma \beta)}{\sqrt{2a\frac{I_0(0)}{Q^4(\beta)}}}.$$
Function $\mathcal{D}$ was defined in Eq. (\ref{afindim3}). Using the Poincar\' e - type
expansion (\ref{poincare}) for function $\mathcal{D}_{-1/2}(z)$ we have for the second part of Fourier transformation
$$\int_{-\infty}^{\infty}d\beta\ e^{-iE\beta}$$
 of the formula (\ref{prp}) the result:
\begin {eqnarray}
& &\Gamma\left(1/4 + \frac{i E}{2 \gamma}\right)\Gamma\left(1/4 - \frac{i E}{2 \gamma}\right) -\label{rozvoj2} \\
&-&a\ \pochh{1/2}{2}\ \int_{-\infty}^{\infty}d\beta\ \frac{e^{-iE\beta}}{\sqrt{\cosh(\gamma \beta)}}\ \frac{3\gamma\beta+\cosh(\gamma\beta)\sinh(\gamma\beta)\left(-4+\cosh^2(\gamma\beta)+\sinh^2(\gamma\beta)\right)}
    {c^2 \gamma^3 \sinh^2(\gamma\beta) \cosh^2(\gamma\beta)}+\cdots \nonumber
    \end {eqnarray}

When we compare Eqs. (\ref{rozvoj}) and (\ref{rozvoj2}) we are capable to evaluate the shift of energy levels due to an-harmonic term.
 Let us stress the term $a \pochh{1/2}{2}=3a/4$ in front of integral. It is known from
the textbooks that the first perturbative correction to energy levels of harmonics oscillator is proportional to term $3a/4.$ In previous qualitative evaluation this term appeared owe to first part of Fourier transformation leading to parabolic cylinder function $D_{-1/2}(z),$
because the Poincar\' e - type expansion of this function possesses this factor in first non-trivial term.

\appendix

\section{ Evaluation of the $N-$ dimensional integral}

Our aim is to evaluate the finite dimensional integral by virtue of the time slicing method:

\begin {equation}
\mathcal{W}_{N}=
\left(\frac{1}{\sqrt{\frac{2\pi\triangle}{c}}}\right)^N
 \int\limits
_{-\infty}^{+\infty} \prod \limits _{i=1}^{max} d\varphi_i
\exp(-L_{N}) , \label{findim}
\end {equation}
where
\begin {equation}
L_{N}= \sum\limits _{i=1}^N \triangle\left[c/2
\left(\frac{\varphi_i-\varphi_{i-1}}{\triangle}\right)^2
+b\varphi_i^2+a\varphi_i^4\right], \label{findim1}
\end {equation}
The dimension $max$ of the integral is equal to $N$ for the
path integral with unconditional Wiener measure, and for the
path integral with conditional Wiener measure $max$ is $N-1$.
The value $\triangle = \beta/N$ correspond to the partition of the imaginary time interval $<0, \beta>$
and in the continuum limit $N\rightarrow \infty$ is going to zero.
The action $L_{N}$ from point of view of the integration variables can
be rewritten in the form:

\begin {eqnarray}
L_N& = & -\frac{c}{\triangle}\varphi_0 \varphi_1 + \frac{c}{2
\triangle}\varphi_0^2 \nonumber \\
& + & \triangle a \varphi_1^4\ + \ (\frac{c}{\triangle}+\triangle
b)\ \varphi_1^2\ - \
\frac{c}{\triangle} \varphi_1 \varphi_2 \nonumber \\
& + & \cdots \nonumber\\
& + & \triangle a \varphi_i^4\ + \ (\frac{c}{\triangle}+\triangle
b)\  \varphi_i^2\ - \ \frac{c}{\triangle} \varphi_i
\varphi_{i+1} \\
& + & \cdots \nonumber\\
& + & \triangle a\varphi_N^4\ + \ (\frac{c}{2 \triangle}+\triangle
b)\ \varphi_N^2\ . \nonumber
\end {eqnarray}

Inserting $L_{N}$ to $\mathcal{W}_{N}$ and by Taylor's expansions of
the terms linear in the integration variables:
$$ exp\{ \frac{c}{\triangle}\varphi_i
\varphi_{i+1}\} =
\sum\limits_{n_i=0}^{\infty}\frac{\left(\sqrt{c/\triangle}\
\varphi_i\ \sqrt{c/\triangle}\ \varphi_{i+1}\right)^{n_i}}{n_i !}\
,$$

we find for the $N-1$ dimensional conditional Wiener measure
integral:

\begin {eqnarray}
\mathcal{W}_{N}&=&\left(\frac{1}{\sqrt{\frac{2\pi\triangle}{c}}}\right)^N
\sum\limits_{n_0=0}^{\infty}\sum\limits_{n_1=0}^{\infty} \cdots
\sum\limits_{n_{N-1}=0}^{\infty}\
\frac{\left(\sqrt{c/\triangle}\ \varphi_0\right)^{n_0}}{n_0 !}\exp{\left\{-\frac{c}{2\triangle}\varphi_0^2\right\}}\nonumber \\
& &
\int\limits_{-\infty}^{+\infty}d\varphi_1\frac{\left(\sqrt{c/\triangle}\
\varphi_1\right)^{n_0+n_1}}{n_1 !}\exp{\left\{-a\triangle \varphi_1^4 -(\frac{c}{\triangle}+b\triangle)\varphi_1^2\right\}}\nonumber \\
& & \cdots \nonumber \\
& &
\int\limits_{-\infty}^{+\infty}d\varphi_i\frac{\left(\sqrt{c/\triangle}\
\varphi_i\right)^{n_{i-1}+n_i}}{n_i !}\exp{\left\{-a\triangle \varphi_i^4 -(\frac{c}{\triangle}+b\triangle)\varphi_i^2\right\}}\nonumber \\
& & \cdots \label{findim2} \\
& &\int\limits
_{-\infty}^{+\infty}d\varphi_{N-1}\frac{\left(\sqrt{c/\triangle}\
\varphi_{N-1}\right)^{n_{N-2}+n_{N-1}}}{n_{N-1} !}\exp{\left\{-a\triangle \varphi_{N-1}^4 -(\frac{c}{\triangle}+b\triangle)\varphi_{N-1}^2\right\}}\nonumber \\
& & \left(\sqrt{c/\triangle}\
\varphi_{N}\right)^{n_{N-1}}\exp{\left\{-a\triangle \varphi_{N}^4
-(\frac{c}{2\triangle}+b\triangle)\varphi_{N}^2\right\}}\nonumber
\end {eqnarray}
In the case of the unconditional Wiener measure integral, in the
last row the operator
$$\int\limits_{-\infty}^{+\infty}d\varphi_{N}$$
will appear and $\mathcal{W}_{N}$ will be the $N$ dimensional
integral. From the above equation is obvious, that for conditional
measure integral the nonzero contribution we obtain only if the
powers of the integration variables are even. This means that all
summation indices $n_i$ must be even, or odd. In the general case we
will evaluate two contributions for the summation indices even and
another for the summation indices odd. For the unconditional measure
integral the situation is simpler, because due to the $d\varphi_{N}$
integration and the power term $\varphi_{N}^{n_{N-1}}$ the nonzero
contribution we find for the even summation indices only.

In this article we evaluate the conditional measure integral with
path starting in the point $\varphi_0 = 0.$ In such case the nonzero
contribution appear for the index $n_0 = 0$ only and due to the
integration over $d\varphi_1$ the nonzero contribution will be obtained
for even summation indices only.

We express the integrals in relation for $\mathcal{W}_{N}$ by
parabolic cylinder functions \cite{prud} and for even summation
indices $n_i=2k_i$ we have:
\begin {equation}
\int\limits_{-\infty}^{+\infty}d\varphi_i\frac{\left(c/\triangle\
\varphi_i^2\right)^{k_{i-1}+k_i}}{2k_i !}\exp{\left\{-a\triangle
\varphi_i^4 -(\frac{c}{\triangle}+b\triangle)\varphi_i^2 \right\}}=
\end {equation}
$$=\left(c/\triangle\ (1+\frac{b\triangle^2}{c})\right)^{-1/2}
\left[c/\triangle\
(1+\frac{b\triangle^2}{c})\right]^{-k_{i-1}-k_i}\Gamma(k_{i-1}+k_i+1/2)\mathcal{D}_{-k_{i-1}-k_i-1/2}(z)$$
Here we have defined:
$$\mathcal{D}_{-k_{i-1}-k_i-1/2}(z) = z^{k_{i-1}+k_i+1/2}\ \exp{\left\{\frac{z^2}{4}\right\}}D_{-k_{i-1}-k_i-1/2}(z)$$
and
$$z = \frac{c/\triangle\left(1+\frac{b\triangle^2}{c}\right)}{\sqrt{2a\triangle}}$$
$D_{\nu}(z)$ is function of the parabolic cylinder of the argument
$z$ and index $\nu$.

Now, inserting all to the relation for $\mathcal{W}_{N}$, we have
for $N-1$ dimensional conditional Wiener measure integral with
starting point $\varphi_0 = 0$ the exact relation without any
approximations:

\begin {eqnarray}
\mathcal{W}_{N}& = &
\left(\frac{2\pi\triangle}{c}\right)^{-1/2}\left[\sqrt{2\pi(1+\frac{b\triangle^2}{c})}\right]^{-N+1}
\sum\limits_{k_1=0}^{\infty}\sum\limits_{k_2=0}^{\infty} \cdots
\sum\limits_{k_{N-1}=0}^{\infty}\nonumber \\
& &
\prod\limits_{i=1}^{N-2}\left\{\frac{(1+\frac{b\triangle^2}{c})^{-2k_i}}{(2k_i)!}
\Gamma(k_{i-1}+k_i+1/2)\mathcal{D}_{-k_{i-1}-k_i-1/2}(z)\right\}\label{findim3}\\
& &
\frac{(1+\frac{b\triangle^2}{c})^{-k_{N-1}}}{(2k_{N-1})!}\left(\frac{c}{\triangle}
\varphi_N^2\right)^{k_{N-1}}\Gamma(k_{N-2}+k_{N-1}+1/2)\mathcal{D}_{-k_{N-2}-k_{N-1}-1/2}(z)\nonumber
\\
& &\exp{\left\{-a\triangle\varphi_N^4 -
(\frac{c}{2\triangle}+b\triangle)\varphi_N^2\right\}}\nonumber
\end {eqnarray}

\section{ Evaluation of the $\mathcal{W}_{N}$ by the recurrent summation over $k_i$ indices}

To perform the summations over indexes $k_i$ in Eq. (\ref{findim3}) we use the
identities:

$$(2k_i)! = 2^{2k_i}\ k_i!\ \pochh{1/2}{k_i}$$
and $$\Gamma(k_{i-1}+k_i+1/2) = \Gamma(1/2)\ \pochh{1/2}{k_{i-1}}\
\pochh{k_{i-1}+1/2}{k_i}$$

Then Eq. (\ref{findim3}) will be rewritten as:

\begin {eqnarray}
\mathcal{W}_{N}& = &
\left(\frac{2\pi\triangle}{c}\right)^{-1/2}\left[\sqrt{2(1+\frac{b\triangle^2}{c})}\right]^{-N+1}
\nonumber \\
& & \prod\limits_{i=1}^{N-2}\left\{\sum\limits_{k_i=0}^{\infty}
\frac{\left[2(1+\frac{b\triangle^2}{c})\right]^{-2k_i}}{k_i!}
\pochh{k_{i-1}+1/2}{k_i}\mathcal{D}_{-k_{i-1}-k_i-1/2}(z)\right\}\label{recur1}\\
& & \sum\limits_{k_{N-1}=0}^{\infty}
\frac{\left[4(1+\frac{b\triangle^2}{c})\right]^{-k_{N-1}}}{(k_{N-1})!\
\pochh{1/2}{k_{N-1}}}\left(\frac{c}{\triangle}
\varphi_N^2\right)^{k_{N-1}}\pochh{k_{N-2}+1/2}{k_{N-1}}\mathcal{D}_{-k_{N-2}-k_{N-1}-1/2}(z)\nonumber
\\
& &\exp{\left\{-a\triangle\varphi_N^4 -
(\frac{c}{2\triangle}+b\triangle)\varphi_N^2\right\}}\nonumber
\end {eqnarray}
As the technical point we turn attention to the same number of the terms $\sqrt{2(1+\frac{b\triangle^2}{c})}$ as the numbers
of the summations over $k_i .$ Therefore we will consider this term with each summation procedure.

We will use the summation relation for parabolic cylinder function
\cite{bateman}
\begin {equation}
e^{x^2/4}\sum\limits_{k=0}^{\infty}\;
\frac{\pochh{\nu}k}{k!}\;t^k\;D_{-\nu-k}(x)\;=\;e^{(x-t)^2/4}\;D_{-\nu}\;(x-t)
\end {equation}
For the functions $\mathcal{D}_{-\nu-k}(x), $ this summation can be
read:
\begin {equation}
\sum\limits_{k=0}^{\infty}\; \frac{\pochh{\nu}k}{k!}\;t^k\;
\mathcal{D}_{-\nu-k}(x)\;= \
\left(\frac{x}{x-xt}\right)^{\nu}\mathcal{D}_{-\nu}(x-xt)\label{recur11}
\end {equation}

The direct application of this identity to Eq. (\ref{recur1}) is not
possible because each function $\mathcal{D}_{-\nu-k}(x)$ possesses
two summation indices and de facto we must evaluate the sum:
\begin {equation}
\frac{1}{\sqrt{2(1+\frac{b\triangle^2}{c})}}\sum\limits_{k_1=0}^{\infty}
\frac{\left[2(1+\frac{b\triangle^2}{c})\right]^{-2k_1}}{k_1!}
\pochh{1/2}{k_1}\mathcal{D}_{-k_1-1/2}(z)
\pochh{k_{1}+1/2}{k_2}\mathcal{D}_{-k_{1}-k_2-1/2}(z)\label{recur2}
\end {equation}
Let us remember the identity:
$$\pochh{1/2}{k_1}\ \pochh{k_{1}+1/2}{k_2} =
\pochh{1/2}{k_2+k_1}\ .$$ We show in the previous article \cite{my}
that sum in Eq. (\ref{recur2}) is uniformly convergent, therefore by
replacing the infinite summation by finite one we can approximate
the infinite sum by the desirable precision. In the finite sum of
the type Eq.(\ref{recur2})
 we can adopt the Poincar\' e-type
 expansion of the parabolic cylinder function, which means:
 \begin {equation}
\mathcal{D}_{-k_1-1/2}(z)\; \equiv \;
z^{k_1+1/2}\;e^{z^2/4}\;D_{-k_1-1/2}(z)\;=
\;\sum\limits_{j=0}^{\mathcal{J}}\; (-1)^j
\;\frac{\pochh{k_1+1/2}{2j}}{j!\;(2z^2)^j}
+\varepsilon_{\mathcal{J}}(k_1, z)
\end {equation}
In the last relation, $\mathcal{J}$ denotes the number of  terms of
the asymptotic expansions convenient to take into account,
$\varepsilon_{\mathcal{J}}(k_1, z)$ is the remainder. We have
discussed the problem of this remainder in our previous
article\cite{my} where we shown that it converge to zero stronger
than $1/N.$ This means, that all contributions to the summations
over indices $k_i$ containing such remainder, or products of
remainders disappears in continuum limit. Our evaluations and
estimates concerning upper limit of this remainder follows from
works of Olver\cite{olver}, Vidunas and Temme\cite{temme2} concerned
to estimates of the upper bounds of remainders of the Poincar\' e
type expansions of the parabolic cylinder functions.

We apply the Poincar\' e type asymptotic expansion for the function
$\mathcal{D}_{-k_{1}-1/2}\;(z)$ and for the leading term (i.e. the term without remainder) of the sum
of the finite set which approximate the Eq. (\ref{recur2}) we have:
\begin {equation}
\frac{1}{\sqrt{2(1+\frac{b\triangle^2}{c})}}\sum\limits_{k_1=0}^{M}
\frac{\left[2(1+\frac{b\triangle^2}{c})\right]^{-2k_1}}{k_1!}
\left\{\sum\limits_{j=0}^{\mathcal{J}}\; (-1)^j
\;\frac{\pochh{k_1+1/2}{2j}}{j!\;(2z^2)^j}\right\}
\pochh{1/2}{k_{2}+k_1}\mathcal{D}_{-k_{1}-k_2-1/2}(z)\label{recur3}
\end {equation}
By replacing the order of summations we read:
\begin {equation}
\frac{1}{\sqrt{2(1+\frac{b\triangle^2}{c})}}\sum\limits_{j=0}^{\mathcal{J}}\; (-1)^j \;\frac{1}{j!\;(2z^2)^j}
\sum\limits_{k_1=0}^{M}
\frac{\left[2(1+\frac{b\triangle^2}{c})\right]^{-2k_1}}{k_1!}
\pochh{k_1+1/2}{2j}
\pochh{1/2}{k_{2}+k_1}\mathcal{D}_{-k_{1}-k_2-1/2}(z)\label{recur4}
\end {equation}
It was proven in previous article \cite{my} the relation:
\begin {equation}
\pochh{k_{1}+1/2}{2j}\; = \;
\sum\limits_{i=0}^{\min{(2j,k_{1})}}a_i^{2j}\frac{(k_{1})!}{(k_{1}-i)!}
\label{recur41}
\end {equation}
Where the coefficients $a_i^{2j}$ are given by:
\begin {equation}
a_i^{j}\; = \;\binom{j}{i}\frac{\pochh{1/2}{j}}{\pochh{1/2}{i}}
\label{recur42}
\end {equation}
We insert these relations into Eq. (\ref{recur4}) and we find:
\begin {equation}
\frac{1}{\sqrt{2(1+\frac{b\triangle^2}{c})}}\sum\limits_{j=0}^{\mathcal{J}}\; (-1)^j \;\frac{1}{j!\;(2z^2)^j}
\sum\limits_{i=0}^{2j}a_i^{2j} \sum\limits_{k_1=i}^{M}
\frac{\left[2(1+\frac{b\triangle^2}{c})\right]^{-2k_1}}{(k_1-i)!}
\pochh{1/2}{k_{2}+k_1}\mathcal{D}_{-k_{1}-k_2-1/2}(z)\label{recur5}
\end {equation}
The series in the summation over $k_i$ is uniformly convergent, in
the precision desired we can extend the summation to infinity. If we
define
$$\sigma = \left[2(1+\frac{b\triangle^2}{c})\right]^{-1}$$
we find:
\begin {equation}
\frac{1}{\sqrt{2(1+\frac{b\triangle^2}{c})}}\sum\limits_{j=0}^{\mathcal{J}}\; (-1)^j \;\frac{1}{j!\;(2z^2)^j}
\sum\limits_{i=0}^{2j}a_i^{2j}\pochh{1/2}{k_{2}+i}\ \sigma^{2i}
\sum\limits_{k_1=i}^{\infty} \frac{\sigma^{2k_1-2i}}{(k_1-i)!}
\pochh{k_{2}+i+1/2}{k_1-i}\mathcal{D}_{-(k_{1}-i)-(k_2+i)-1/2}(z)\label{recur6}
\end {equation}
By identity  Eq.(\ref{recur11}) we have for the leading term of Eq. (\ref{recur2}):
\begin {equation}
\frac{1}{\sqrt{2(1+\frac{b\triangle^2}{c})}}\sum\limits_{j=0}^{\mathcal{J}}\; (-1)^j \;\frac{1}{j!\;(2z^2)^j}
\sum\limits_{i=0}^{2j}a_i^{2j}\ \sigma^{2i}
\left(\frac{z}{z-z\sigma^2}\right)^{k_{2}+i+1/2}\
\pochh{1/2}{k_{2}+i}\
\mathcal{D}_{-k_{2}-i-1/2}(z-z\sigma^2)\label{recur7}
\end {equation}
In this evaluation we have neglected the remainders of two uniformly convergent series. Such replacement
can be done with desired precision. The necessity of the truncation of the series depends on the dimension
of the integral therefore in continuum limit the truncated series approach to the original series.
For the recurrence procedure we define the new variables:
$$\sigma_1 = \sigma^2$$
$$z_1 = z(1-\sigma^2)$$
$$\omega_1 = \frac{z_1}{z} = 1-\sigma_1$$
$$(1)^{2j}_i = a^{2j}_i$$

We follow with next recurrence step, the summation over index $k_2.$
As we see from Eqs. (\ref{recur1}) and (\ref{recur7}), we must
provide the sum:
\begin {eqnarray}
& &\frac{1}{\sqrt{2 \omega_1 (1+\frac{b\triangle^2}{c})}}\sum\limits_{j=0}^{\mathcal{J}}\; (-1)^j
\;\frac{1}{j!\;(2z^2)^j} \sum\limits_{i=0}^{2j}\
(1)^{2j}_i\ \left(\frac{\sigma_1}{1-\sigma_1}\right)^i\label{recur8}\\
& & \frac{1}{\sqrt{2(1+\frac{b\triangle^2}{c})}} \sum\limits_{k_2=0}^{\infty}\ \frac{1}{(k_2)!}\
\left(\sigma^2\frac{z}{z_1}\right)^{k_{2}}\ \pochh{1/2}{k_{2}+i}\
\mathcal{D}_{-k_{2}-i-1/2}(z_1)
\pochh{k_{2}+1/2}{k_3}\mathcal{D}_{-k_{2}-k_3-1/2}(z)\nonumber
\end {eqnarray}
By the identity:
$$\pochh{1/2}{k_{2}+i}\ =\ \pochh{1/2}{k_{2}}\ \pochh{k_{2}+1/2}{i}$$
and the leading term for Poincar\' e expansion of the term
$$\pochh{k_{2}+1/2}{i}\  \mathcal{D}_{-k_{2}-i-1/2}(z_1)\ =\
\sum\limits_{j_1=0}^{\mathcal{J}}\; (-1)^{j_1}
\;\frac{\pochh{k_2+1/2}{2j_1+i}}{j_1!\;(2z_1^2)^{j_1}},$$ and with
the new variable

$$\sigma_2 = \sigma^2\frac{z}{z_1}$$

we find:
\begin {eqnarray}
& &\frac{1}{\sqrt{2 \omega_1(1+\frac{b\triangle^2}{c})}}\frac{1}{\sqrt{2(1+\frac{b\triangle^2}{c})}}\sum\limits_{j=0}^{\mathcal{J}}\;
\sum\limits_{j_1=0}^{\mathcal{J}}\; (-1)^{j+j_1}
\;\frac{1}{j!j_1!\;(2z^2)^{j+j_1}}\left(\frac{1}{\omega_1}\right)^{2j_1}
\sum\limits_{i=0}^{2j}\
(1)^{2j}_i\ \left(\frac{\sigma_1}{1-\sigma_1}\right)^i\label{recur9}\\
& & \sum\limits_{k_2=0}^{\infty}\
\frac{\pochh{k_2+1/2}{2j_1+i}}{(k_2)!}\
\left(\sigma_2\right)^{k_{2}}\
\pochh{1/2}{k_{2}+k_3}\mathcal{D}_{-k_{2}-k_3-1/2}(z)\nonumber
\end {eqnarray}

We re-define the summation
$$\sum\limits_{j=0}^{\mathcal{J}}\;
\sum\limits_{j_1=0}^{\mathcal{J}}\; =\
\sum\limits_{\mu=0}^{2\mathcal{J}}\sum\limits_{j_1=0}^{\mu}\
\binom{\mu}{j_1},$$ where $\mu = j+j_1.$ By replacing
$\pochh{k_2+1/2}{2j_1+i}$ by the identity (\ref{recur41}):
$$\pochh{k_2+1/2}{2j_1+i}\ =\
\sum\limits_{i_1=0}^{\min{(2j_1+i,k_{2})}}a_{i_1}^{2j_1+i}\frac{(k_{2})!}{(k_{2}-i_1)!}
$$

we have:
\begin {eqnarray}
& &\frac{1}{\sqrt{2 \omega_1(1+\frac{b\triangle^2}{c})}}\frac{1}{\sqrt{2(1+\frac{b\triangle^2}{c})}}\sum\limits_{\mu=0}^{\mathcal{J}}\;(-1)^{\mu}
\;\frac{1}{\mu!\;(2z^2)^{\mu}} \sum\limits_{j_1=0}^{\mu}\;
\binom{\mu}{j_1}\left(\frac{1}{\omega_1}\right)^{2j_1}
\sum\limits_{i=0}^{2\mu-2j_1}\
(1)^{2\mu-2j_1}_i\ \left(\frac{\sigma_1}{1-\sigma_1}\right)^i\label{recur10}\\
& & \sum\limits_{i_1=0}^{2j_1+i}\ a_{i_1}^{2j_1+i}\
\left(\sigma_2\right)^{i_1} \sum\limits_{k_2=i_1}^{\infty}\
\frac{\left(\sigma_2\right)^{k_{2}-i_1}}{(k_2-i_1)!}\ \
\pochh{1/2}{k_{2}+k_3}\mathcal{D}_{-k_{2}-k_3-1/2}(z)\nonumber
\end {eqnarray}

In the above sum the summation index $i$ is an inner index, we
replace the summations over $i$ and $i_1$
$$\sum\limits_{i=0}^{2\mu-2j_1}\ \sum\limits_{i_1=0}^{2j_1+i}\ =\
\sum\limits_{i_1=0}^{2\mu}\ \sum\limits_{i=\max{(0,\
i_1-2j_1)}}^{2\mu-2j_1}$$

It is obvious that now summations over indices $j_1$ and $i_1$ are
independent and therefore can be replaced. For Eq. (\ref{recur10})
we have:
\begin {eqnarray}
& &\frac{1}{\sqrt{2 \omega_1(1+\frac{b\triangle^2}{c})}}\frac{1}{\sqrt{2(1+\frac{b\triangle^2}{c})}}\sum\limits_{\mu=0}^{\mathcal{J}}\;(-1)^{\mu}
\;\frac{1}{\mu!\;(2z^2)^{\mu}} \sum\limits_{i_1=0}^{2\mu}\
\left(\sigma_2\right)^{i_1} \sum\limits_{j_1=0}^{\mu}\;
\binom{\mu}{j_1}\left(\frac{1}{\omega_1}\right)^{2j_1}\label{recur20}\\
& &\sum\limits_{i=\max{(0,\ i_1-2j_1)}}^{2\mu-2j_1}\
a_{i_1}^{2j_1+i}\ (1)^{2\mu-2j_1}_i\ \left(\frac{\sigma_1}{1-\sigma_1}\right)^i
\sum\limits_{k_2=i_1}^{\infty}\
\frac{\left(\sigma_2\right)^{k_{2}-i_1}}{(k_2-i_1)!}\ \
\pochh{1/2}{k_{2}+k_3}\mathcal{D}_{-k_{2}-k_3-1/2}(z)\nonumber
\end {eqnarray}

The sum over index $k_2$ can be evaluated:
$$\sum\limits_{k_2=i_1}^{\infty}\
\frac{\left(\sigma_2\right)^{k_{2}-i_1}}{(k_2-i_1)!}\ \
\pochh{1/2}{k_{2}+k_3}\mathcal{D}_{-k_{2}-k_3-1/2}(z) =
\pochh{1/2}{k_{3}+i_1}\
\left(\frac{z}{z-z\sigma_2}\right)^{k_{3}+i_1+1/2}\mathcal{D}_{-k_3-i_1-1/2}(z-z\sigma_2)$$

By definitions $$z_2 = z-z\sigma_2\ ,$$ $$\omega_2 = z_2/z\ ,$$ and
$$\left(2\right)_{i_1}^{2\mu} =  \sum\limits_{j_1=0}^{\mu}\;
\binom{\mu}{j_1}\left(\frac{1}{\omega_1}\right)^{2j_1}
\sum\limits_{i=\max{(0,\ i_1-2j_1)}}^{2\mu-2j_1}\ a_{i_1}^{2j_1+i}\
(1)^{2\mu-2j_1}_i\left(\frac{\sigma_1}{1-\sigma_1}\right)^i $$

we find for the leading term of Eq. (\ref{recur20}) the expression:
\begin {equation}
\frac{1}{\sqrt{\prod
\limits_{i=1}^{2}\left(2\omega_i(1+b\triangle^2/c)\right)}}\sum\limits_{\mu=0}^{\mathcal{J}}\;(-1)^{\mu}
\;\frac{1}{\mu!\;(2z^2)^{\mu}}\sum\limits_{i_1=0}^{2\mu}\
\left(2\right)_{i_1}^{2\mu}\left(\frac{\sigma_2}{1-\sigma_2}\right)^{i_1}\pochh{1/2}{k_{3}+i_1}\
\left(\frac{z}{z_2}\right)^{k_{3}}\mathcal{D}_{-k_3-i_1-1/2}(z_2)
\end {equation}

By the method applied to the first two summations we can by the
mathematical induction prove the lemma:

\noindent
\textit{Lemma.} \textit{The leading term of the partial sum of} Eq. (\ref{recur1}) \textit{over the
indices} $k_1,\ k_2,\ \cdots\ ,\ k_{\Lambda}$, $\Lambda \leq N-2$ \textit{is
 the relation:}
\begin {eqnarray}
& & Z_{\Lambda}=\frac{1}{\sqrt{\frac{2\pi\triangle}{c} \prod
\limits_{i=1}^{\Lambda}\left(2\omega_i(1+b\triangle^2/c)\right)}}\\
& & \sum\limits_{\mu=0}^{\mathcal{J}}\ (-1)^{\mu}
\;\frac{1}{\mu!\;(2z^2)^{\mu}}\sum\limits_{i_{\Lambda}=0}^{2\mu}\
\left(\Lambda\right)^{2\mu}_{i_{\Lambda}}\left(\frac{\sigma_{\Lambda}}{1-\sigma_{\Lambda}}\right)^{i_{\Lambda}}
\left(\frac{z}{z_{\Lambda}}\right)^{k_{\Lambda+1}}
\pochh{1/2}{k_{\Lambda+1}+i_{\Lambda}}\
\mathcal{D}_{-k_{\Lambda+1}-i_{\Lambda}-1/2}(z_{\Lambda}).\nonumber
\end {eqnarray}
The symbol $(\Lambda)^{\nu}_{\mu}$ is defined by the recurrence relation:
\begin {equation}
\left(\Lambda\right)^{2\mu}_{i_{\Lambda}}=
\sum\limits_{j_{\Lambda}=0}^{\mu}\;
\binom{\mu}{j_{\Lambda}}\left(\frac{1}{\omega_{\Lambda-1}}\right)^{2j_{\Lambda}}
\sum\limits_{i=\max{(0,\
i_{\Lambda}-2j_{\Lambda}})}^{2\mu-2j_{\Lambda}}\
a_{i_{\Lambda}}^{2j_{\Lambda}+i}\ (\Lambda-1)^{2\mu-2j_{\Lambda}}_i
\left(\frac{\sigma_{\Lambda-1}}{1-\sigma_{\Lambda-1}}\right)^i
\end {equation}
The first term of the recurrence relation is: $$\left(1\right)^{2\mu}_{i}= a^{2\mu}_{i}\ .$$

In the recurrent relation are the another recurrence definition:
$$\sigma_{i+1} = \frac{\sigma^2}{1-\sigma_i}\ ,\ \sigma_1=\sigma^2\
,$$
$$z_i = z(1-\sigma_i)\ ,$$
$$\omega_{i} = 1-\sigma_i$$
or $$\omega_{i+1}=1-\frac{\sigma^2}{\omega_{i}}\ ,\
\omega_1=1-\sigma^2\ , \omega_0=1.$$

To complete the evaluation of the leading term in Eq. (\ref{recur1}),
we must to perform the sum over the last index. This sum  differs from previous
recurrence summations, we have:
\begin {eqnarray}
& &\mathcal{W}_{N} = \frac{1}{\sqrt{2(1+\frac{b\triangle^2}{c})}}\  \sum\limits_{k_{N-1}=0}^{\infty}
\frac{\left[4(1+\frac{b\triangle^2}{c})\right]^{-k_{N-1}}}{(k_{N-1})!\
\pochh{1/2}{k_{N-1}}}\left(\frac{c}{\triangle}
\varphi_N^2\right)^{k_{N-1}}\ Z_{N-2}\nonumber
\\
& &\exp{\left\{-a\triangle\varphi_N^4 -
(\frac{c}{2\triangle}+b\triangle)\varphi_N^2\right\}}\label{recur100}
\end {eqnarray}
Replacing $Z_{N-2}$ by corresponding leading term and by definition $\omega_0=1$, we find:
\begin {eqnarray}
\mathcal{W}_{N} =
\frac{1}{\sqrt{\left(\frac{2\pi\triangle}{c}\right) \prod
\limits_{i=0}^{N-2}2\omega_i(1+b\triangle^2/c)}}\exp{\left\{-a\triangle\varphi_N^4
- (\frac{c}{2\triangle}+b\triangle)\varphi_N^2\right\}}
\label{recur101}
\\
\sum\limits_{\mu=0}^{\mathcal{J}}\ (-1)^{\mu}
\;\frac{1}{\mu!\;(2z^2)^{\mu}}\sum\limits_{i=0}^{2\mu}\
\left(N-2\right)^{2\mu}_{i}\left(\frac{\sigma_{N-2}}{1-\sigma_{N-2}}\right)^i\nonumber\\
\sum\limits_{k_{N-1}=0}^{\infty}\frac{\xi^{k_{N-1}}}{(k_{N-1})!\pochh{1/2}{k_{N-1}}}
\pochh{1/2}{k_{N-1}+i}\mathcal{D}_{-k_{N-1}-i-1/2}(z_{N-2})\nonumber
\end {eqnarray}
where we defined: $$\xi = \frac{1}{\omega_{N-2}}\
\frac{c}{4\triangle(1+\frac{b\triangle^2}{c})}\ \varphi_N^2$$

The uniformly convergent sum
\begin {equation}
\sum\limits_{k_{N-1}=0}^{\infty}\frac{\xi^{k_{N-1}}}{(k_{N-1})!\pochh{1/2}{k_{N-1}}}
\pochh{1/2}{k_{N-1}+i}\mathcal{D}_{-k_{N-1}-i-1/2}(z_{N-2})
\label{app1}
\end {equation}

is evaluated by leading term of  Poincar\' e expansion of the
relation:
$$\pochh{1/2}{k_{N-1}+i}\mathcal{D}_{-k_{N-1}-i-1/2}(z_{N-2})\ =\
\sum\limits_{j=0}^{\mathcal{J}}(-1)^{j}
\;\frac{1}{j!\;(2z_{N-2}^2)^{j}}\pochh{1/2}{k_{N-1}+i+2j}$$
Following the identity:
$$\pochh{1/2}{k_{N-1}+i+2j}\ =\ \pochh{1/2}{i+2j}\ \pochh{1/2+i+2j}{k_{N-1}}$$
we have for the leading part of (\ref{app1}):
$$\sum\limits_{j=0}^{\mathcal{J}}(-1)^{j}
\;\frac{1}{j!\;(2z^2)^{j}}\ \left(\frac{z}{z_{N-2}}\right)^{2j}\
\pochh{1/2}{i+2j}\
\sum\limits_{k_{N-1}=0}^{\infty}\frac{\xi^{k_{N-1}}}{(k_{N-1})!\
\pochh{1/2}{k_{N-1}}}\ \pochh{1/2+i+2j}{k_{N-1}}$$ The sum over
index $k_{N-1}$  can be evaluated \cite{prud} and we have:
$$\pochh{1/2}{i+2j}\
\sum\limits_{k_{N-1}=0}^{\infty}\frac{\xi^{k_{N-1}}}{(k_{N-1})!\
\pochh{1/2}{k_{N-1}}}\ \pochh{1/2+i+2j}{k_{N-1}}\ = \ (i+2j)!\
\exp{(\xi)}\ \mathcal{P}^{-1/2}_{i+2j}(-\xi)\ ,$$ where
$\mathcal{P}^{\lambda}_{n}$ is the generalized Laguere polynomial.
Moreover, in our case holds the identity\cite{bateman}:
$$(i+2j)!\ \mathcal{P}^{-1/2}_{i+2j}(-\xi)\ =\ \sum \limits_{p=0}^{i+2j}\ a^{i+2j}_p\ (\xi)^p\ ,$$
where $a^{i+2j}_p$ is the symbol defined in previous text. With all
this evaluated relations we find for Eq. (\ref{recur101})
\begin {eqnarray}
\mathcal{W}_{N} =
\frac{1}{\sqrt{\left(\frac{2\pi\triangle}{c}\right) \prod
\limits_{i=0}^{N-2}2\omega_i(1+b\triangle^2/c)}}\exp{\left\{-a\triangle\varphi_N^4
- (\frac{c}{2\triangle}+b\triangle)\varphi_N^2\ +\ \xi\right\}}
\label{recur102}
\\
\sum\limits_{\mu=0}^{\mathcal{J}}\
\sum\limits_{j=0}^{\mathcal{J}}(-1)^{j+\mu} \;\frac{1}{j!\
\mu!\;(2z^2)^{j+\mu}}\ \left(\frac{z}{z_{N-2}}\right)^{2j}
\sum\limits_{i=0}^{2\mu}\ \sum \limits_{p=0}^{i+2j}\
\left(N-2\right)^{2\mu}_{i}\left(\frac{\sigma_{N-2}}{1-\sigma_{N-2}}\right)^i
\ a^{i+2j}_p\ (\xi)^p\nonumber
\end {eqnarray}
In the next steep we handle the summations over $\mu$ and $j$ as:
$$\sum\limits_{\mu=0}^{\mathcal{J}}\
\sum\limits_{j=0}^{\mathcal{J}}(-1)^{j+\mu} \;\frac{1}{j!\
\mu!\;(2z^2)^{j+\mu}}\ =\ \sum\limits_{\nu=0}^{\mathcal{J}}\
(-1)^{\nu} \ \frac{1}{\nu!\;(2z^2)^{\nu}} \sum\limits_{j=0}^{\nu}\
\binom{\nu}{j}\ ,$$ and we replace $\mu = \nu-j\ ,$ as well as we
replace the summations over the indices $i$ and $p\ :$
$$\sum\limits_{i=0}^{2\nu-2j}\ \sum \limits_{p=0}^{i+2j}\
=\ \sum \limits_{p=o}^{2\nu}\ \sum\limits_{i=\max{(0\ ,
p-2j)}}^{2\nu-2j}$$ Then we find for Eq. (\ref{recur102}):
\begin {eqnarray}
\mathcal{W}_{N} &=&
\frac{1}{\sqrt{\left(\frac{2\pi\triangle}{c}\right) \prod
\limits_{i=0}^{N-2}2\omega_i(1+b\triangle^2/c)}}\exp{\left\{-a\triangle\varphi_N^4
- (\frac{c}{2\triangle}+b\triangle)\varphi_N^2\ +\ \xi\right\}}
\sum\limits_{\nu=0}^{\mathcal{J}}\ (-1)^{\nu} \
\frac{1}{\nu!\;(2z^2)^{\nu}}\ \sum \limits_{p=0}^{2\nu}\ (\xi)^p
\nonumber \\
& &\sum\limits_{j=0}^{\nu}\ \binom{\nu}{j}
\left(\frac{1}{\omega_{N-2}}\right)^{2j} \sum\limits_{i=\max{(0\ ,
p-2j)}}^{2\nu-2j}
\left(N-2\right)^{2\nu-2j}_{i}\left(\frac{\sigma_{N-2}}{1-\sigma_{N-2}}\right)^i
\ a^{i+2j}_p \label{recur103}
\end {eqnarray}
Following \textit{Lemma} the expression in the last line of the above equation can be replaced by
 $$\left(N-1\right)^{2\nu}_{p}\ .$$ After summation over
all indices $k_i$ we finally find the leading term to  Eq (\ref{recur1}):
\begin {eqnarray}
\mathcal{W}_{N} &=&
\frac{1}{\sqrt{\left(\frac{2\pi\triangle}{c}\right) \prod
\limits_{i=0}^{N-2}2\omega_i(1+b\triangle^2/c)}}\exp{\left\{-a\triangle\varphi_N^4
- (\frac{c}{2\triangle}+b\triangle)\varphi_N^2\ +\ \xi\right\}}\nonumber \\
& & \sum\limits_{\nu=0}^{\mathcal{J}}\ (-1)^{\nu} \
\frac{1}{\nu!\;(2z^2)^{\nu}}\ \sum \limits_{p=0}^{2\nu}\ (\xi)^p\
\left(N-1\right)^{2\nu}_{p} \label{recur104}
\end {eqnarray}

\noindent
All non-leading terms due to the remainders of the Poincar\' e expansion of the parabolic cylinder function
in the continuum limit disappears \cite{my}.  We will discuss the continuum limit of the Eq.
(\ref{recur104}) in the next Appendices.

\section{ The continuum limit of the square-root factor in the Eq.( \ref{recur104}).}

 In our article
\cite{my} we evaluated the continuum  limit of the leading part of the $N$ dimensional integral $\mathcal{W}_N$
 by generalized Gelfand -- Yaglom
equation. We defined the function $F_N\ ,$ connected with $N$ dimensional integral
by the relation:
$$\mathcal{W}_N = \frac{1}{\sqrt{F_N}}.$$
Due to the recurrence relations for the quantities in $\mathcal{W}_N$ we can evaluate the difference
equation for the values $F_k,$ where $k=1, 2, \cdots , N$
The aim of the Gelfand-Yaglom construction is to find the continuum
limit of the difference equation for the function $F_k.$ Solution of
this differential equation is connected to the continuum path
integral by:
$$\mathcal{W(\beta)} =\frac{1}{\sqrt{F(\beta)}},$$
where $\beta$ is the upper bound of the time interval in the action.

We can use the same method
to evaluate the continuum limit of the $N$ dimensional integral for the conditional measure Wiener integral,
 but we would like to present slightly different method here. In the Gelfand -- Yaglom method, we evaluated the difference equation, after continuum limit we obtain the differential equation and we find its solution. In the new approach, we evaluate directly this function.
We will present the evaluation of the continuum limit of the relation

\begin {equation}
\left(\frac{2\pi\triangle}{c}\right) \prod
\limits_{i=0}^{N-2}2\omega_i(1+b\triangle^2/c)\ ,\label{gy1}
\end {equation}
where $\omega_i$ obeys the recurrence relation:
$$\omega_{i+1} = 1-\frac{\sigma^2}{\omega_i}\ ,$$
where $$\omega_0 = 1\ ,\ \omega_1 = 1-\sigma^2\ ,\ \sigma =
\left[2(1+\frac{b\triangle^2}{c})\right]^{-1}\ .$$

Let us define:
$$\Omega_n =  \prod
\limits_{i=0}^{n}\omega_i\ .$$
By the recurrence relation for $\omega_n$ we have the recurrence relation for $\Omega_n$:
\begin {equation}
\Omega_n = \omega_n\ \omega_{n-1}\ \Omega_{n-2}\ =\ (\omega_{n-1}-\sigma^2)\ \Omega_{n-2}\ =
\ \Omega_{n-1}-\sigma^2\ \Omega_{n-2}\ ,\label{gy2}
\end {equation}
with first two values:
$$\Omega_0 = 1,\ \Omega_1 = 1-\sigma^2\ .$$
The methods of the difference calculus \cite{findif} propose to search for solution of the recurrence
equation (\ref{gy2}) in the form:

\begin {equation}
\Omega_n = w_1\ \varrho_1^n + w_2\ \varrho_2^n
\end {equation}
We find for $\varrho$ the characteristic equation
$$\varrho^2-\varrho + \sigma^2 = 0\ ,$$
with solution
$$\varrho_{1,2} = \frac{1\pm\sqrt{1-4\sigma^2}}{2}$$
The coefficients $w_1\ ,\ w_2$ will be evaluated from the values $\Omega_0\ ,\ \Omega_1$ and we find:
$$w_{1,2} = 1/2\left(1\pm\frac{1-2\sigma^2}{\sqrt{1-4\sigma^2}}\right)\ .$$

For expression (\ref{gy1}) we have:

\begin {equation}
\left(\frac{2\pi\triangle}{c}\right)
\left[2(1+b\triangle^2/c)\right]^{N-1}\ \Omega_{N-2}\ =\
\left(\frac{2\pi\triangle}{c}\right)
\left[2(1+b\triangle^2/c)\right]^{N-1}\ (w_1\ \rho_1^{N-2}\ +\ w_2\
\rho_2^{N-2}) \label{gy3}
\end {equation}
Inserting $w_{1,2}$ and $\rho_{1,2}$ after some algebra we find:

\begin {eqnarray}
\frac{4\pi\triangle}{c}\left(1+\frac{b\triangle^2}{c}\right)
\left\{\frac{1}{2}\left(1+\frac{1-2\sigma^2}{\sqrt{1-4\sigma^2}}\right)
\left[\left(1+\frac{b\triangle^2}{c}\right)\left(1+\sqrt{1-4\sigma^2}\right)
\right]^{N-2}\ + \right. \nonumber\\ \left.
+\frac{1}{2}\left(1-\frac{1-2\sigma^2}{\sqrt{1-4\sigma^2}}\right)
\left[\left(1+\frac{b\triangle^2}{c}\right)\left(1-\sqrt{1-4\sigma^2}\right)
\right]^{N-2}
\right\}\label{gy4}
\end {eqnarray}

Let us define the value
$$\gamma = \sqrt{2b/c}\ ,$$
and by the definition of the symbol
$$\triangle = \frac{\beta}{N}\ ,$$
we find in the continuum limit $\lim{N}\rightarrow \infty$
 for the  expression (\ref{gy4}) the result:
$$\frac{2\pi}{c}\ \frac{\sinh(\gamma \beta)}{\gamma}\ .$$
The same result we find by Gelfand -- Yaglom method also.

\section{ The continuum limit of the exponential factor in the Eq.(\ref{recur104}).}

We are going to evaluate the exponent in Eq. (\ref{recur104}):
\begin {equation}
\exp{\left\{-a\triangle\varphi_N^4 -
(\frac{c}{2\triangle}+b\triangle)\varphi_N^2\ +\ \xi\right\}}
\label{expo01}
\end{equation}
where $\xi$ is defined as:
$$\xi = \frac{1}{\omega_{N-2}}\
\frac{c}{4\triangle(1+\frac{b\triangle^2}{c})}\ \varphi_N^2\ ,$$ and
$\omega_{N-2}$ obey the recurrence relation:
$$\omega_{i+1} = 1-\frac{\sigma^2}{\omega_i}\ ,\ \omega_1 = 1-\sigma^2\ ,\  \omega_2 = \frac{1-2\sigma^2}{1-\sigma^2}$$
We are going to evaluate the $\omega_i.$
Following the method of the $n-th$ convergent \cite{findif} we
define:
$$\omega_n = \frac{p_n}{q_n} = a_n + \frac{b_n}{\omega_{n-1}}\ ,$$
for $p_n$ and $q_n$ we find the equations:
$$p_n = a_n p_{n-1}+b_n p_{n-2}$$
$$q_n = a_n q_{n-1}+b_n q_{n-2}$$
the solutions of the above recurrence equations can be written in
the form:
$$p_n = u_1 \rho_1^n + u_2 \rho_2^n$$
$$q_n = \tilde{u}_1 \rho_1^n + \tilde{u}_2 \rho_2^n$$
Where $\rho$ is solution of the characteristic equation:
$$\rho^2 -a_n \rho -b_n = 0\ ,\ a_n = 1\ ,\ b_n = -\sigma\ .$$
The solution of this equation is:
\begin {equation}
\rho_{1,2} = \frac{1 \pm \sqrt{1-4\sigma^2}}{2}
\label{expo1}
\end {equation}
The coefficients $u_{1,2}$ and $\tilde{u}_{1,2}$ we find from the
conditions:
$$\omega_1 = \frac{p_1}{q_1} = 1-\sigma^2\ ,\ \omega_2 =
\frac{p_2}{q_2} = \frac{1-2\sigma^2}{1-\sigma^2}\ .$$
We find:
\begin {eqnarray}
u_{1,2} = \frac{1}{2}\left(1 \pm
\frac{1-2\sigma^2}{\sqrt{1-4\sigma^2}}\right)\label{expo2}\\
\tilde{u}_{1,2} = \frac{1}{2}\left(1 \pm
\frac{1}{\sqrt{1-4\sigma^2}}\right)\nonumber
\end {eqnarray}
From Eqs. (\ref{expo1},\ref{expo2}) follows the important identity:
$$
q_n = p_{n-1}\ .
$$
This identity allows us to write $\omega_n$ as:
\begin {equation}
\omega_n = \frac{p_n}{p_{n-1}} = \frac{q_{n+1}}{q_n}\ .
\end {equation}
In this point we define the new variables for the evaluations of the continuum limit. Following the definition
of $\omega_n$ we have:
\begin {equation}
\omega_n =  \frac{q_{n+1}}{q_n}\ =\ \frac{\tilde{u}_1 \rho_1^{n+1} + \tilde{u}_2 \rho_2^{n+1}}{\tilde{u}_1 \rho_1^n + \tilde{u}_2 \rho_2^n}\ =\
\sigma \frac{\left(\frac{\rho_1}{\sigma}\right)^{n+1} + \frac{\tilde{u}_2}{\tilde{u}_1} \left(\frac{\rho_2}{\sigma}\right)^{n+1}}{\left(\frac{\rho_1}{\sigma}\right)^{n} + \frac{\tilde{u}_2}{\tilde{u}_1} \left(\frac{\rho_2}{\sigma}\right)^{n}} =\ \sigma\ \frac{Q_{n+1}}{Q_n} .
\label{apd5}
\end {equation}

For the following evaluations we will use the definition of the variable $Q_n$:
\begin {equation}
Q_n\ =\ \left(\frac{\rho_1}{\sigma}\right)^{n} + \frac{\tilde{u}_2}{\tilde{u}_1} \left(\frac{\rho_2}{\sigma}\right)^{n}\ .
\label{expo22}
\end {equation}
Comparing to $q_n$, the variables $Q_n$ are finite in the continuum limit. The primary variable in our calculation is variable
$\omega_n\ ,$ finite in the continuum limit and it is more expedient from point of view of the following evaluation to express
it as the proportion of the $Q_n$. For complexity, we define the variable $\tilde{Q}_n$:
\begin {equation}
\tilde{Q}_n\ =\ \left(\frac{\rho_1}{\sigma}\right)^{n} - \frac{\tilde{u}_2}{\tilde{u}_1} \left(\frac{\rho_2}{\sigma}\right)^{n}\ .
\label{expo23}
\end {equation}

To evaluate the relation (\ref{expo01}) we done the key calculation:
$$\frac{1}{\omega_{N-2}} = \frac{q_{N-1}}{q_{N-2}} =
2\left(1-\frac{\sqrt{1-4\sigma^2}\
(\tilde{u}_1\rho_1^{N-2}-\tilde{u}_2\rho_2^{N-2})}{(\tilde{u}_1\rho_1^{N-2}+\tilde{u}_2\rho_2^{N-2})
+\sqrt{1-4\sigma^2}(\tilde{u}_1\rho_1^{N-2}-\tilde{u}_2\rho_2^{N-2})}\right)$$
Inserting to Eq. (\ref{expo01}) we find the exponent in the form:
\begin {eqnarray}
&-& a\triangle\varphi_N^4 - \varphi_N^2\
\left(\frac{c}{2\triangle}+b\triangle-\frac{c}{2\triangle(1+b\triangle^2/c)}\right)\label{expo3}\\
&-& \left(\frac{\sqrt{1-4\sigma^2}\
(\tilde{u}_1\rho_1^{N-2}-\tilde{u}_2\rho_2^{N-2})}{(\tilde{u}_1\rho_1^{N-2}+\tilde{u}_2\rho_2^{N-2})
+\sqrt{1-4\sigma^2}(\tilde{u}_1\rho_1^{N-2}-\tilde{u}_2\rho_2^{N-2})}\right)
\frac{c}{2\triangle(1+\frac{b\triangle^2}{c})}\varphi_N^2\nonumber
\end {eqnarray}
In continuum limit, where $\lim{N}\rightarrow \infty\ ,$ the terms
in the first row of Eq. (\ref{expo3}) are going to zero. In the
second row, with help of the identities:
$$\sqrt{1-4\sigma^2} = \frac{\triangle\
\sqrt{2b/c+b^2\triangle^2/c^2}}{1+b\triangle^2/c}\ ,$$
$$\lim_{N\rightarrow \infty}\left(\frac{\tilde{u}_2}{\tilde{u}_1}\right) =
\lim_{N\rightarrow
\infty}\left(\frac{\sqrt{1-4\sigma^2}-1}{\sqrt{1-4\sigma^2}+1}\right)
= -1\ ,$$ and
$$\lim_{N\rightarrow \infty}(2\rho_{1,2})^{N-2} =
\lim_{N\rightarrow \infty}(1 \pm \sqrt{1-4\sigma^2})^N =
\lim_{N\rightarrow \infty}\left(1 \pm
\frac{\triangle\sqrt{2b/c+b^2\triangle^2/c^2}}{1+b\triangle^2/c}\right)^N
= \exp{(\pm \gamma t)}\ ,$$ we find for $\lim{N\rightarrow\infty}$
of the Eq. (\ref{expo3}) the result:
\begin {equation}
- \frac{c \gamma}{2}\coth{(\gamma t)\varphi_N^2\ ,}
\end {equation}
because we defined previously that:
$$\triangle = t/N\ ,\ \gamma = \sqrt{2b/c}\ .$$

\section{ The evaluation of the an-harmonicity correction in the Eq.(\ref{recur104}).}

We are going to evaluate the term $\left(N-1\right)^{2\mu}_{p}$ appearing in Eq. (\ref{recur104}) in the sum:
\begin {equation}
\sum \limits_{p=0}^{2\mu}\ (\xi)^p\ \left(N-1\right)^{2\mu}_{p}.
\label{mtc1}
\end {equation}
For evaluation we introduce the recurrence procedure, which is explained in this appendix.

We have previously defined in appendix B the quantity:
$$\xi = \frac{1}{\omega_{N-2}}\
\frac{c}{4\triangle(1+\frac{b\triangle^2}{c})}\ \varphi_N^2\ ,$$
and the recurrence relation:
\begin {equation}
\left(\Lambda\right)^{2\mu}_{p}= \sum\limits_{j=0}^{\mu}\;
\binom{\mu}{j}\left(\frac{1}{\omega_{\Lambda-1}}\right)^{2j}
\sum\limits_{i=\max{(0,\ p-2j})}^{2\mu-2j}\ a_{p}^{2j+i}\
(\Lambda-1)^{2\mu-2j}_i
\left(\frac{\sigma_{\Lambda-1}}{1-\sigma_{\Lambda-1}}\right)^i
\label{mtc2}
\end {equation}
when $$\left(1\right)^{2\mu}_{i}= a^{2\mu}_{i}\ .$$

In the recurrent relation we use the another recurrence definition:
$$\sigma_{i+1} = \frac{\sigma^2}{1-\sigma_i}\ ,\ \sigma_1=\sigma^2\
,$$
$$\sigma = \frac{1}{2(1+\frac{b\triangle^2}{c})},$$
$$\omega_{i} = 1-\sigma_i$$
or $$\omega_{i+1}=1-\frac{\sigma^2}{\omega_{i}}\ ,\
\omega_1=1-\sigma^2$$

\noindent
 Following the above relations, we can replace in Eq. (\ref{mtc2})
 the term:
 $$\frac{\sigma_{\Lambda-1}}{1-\sigma_{\Lambda-1}} = \frac{\sigma^2}{\omega_{\Lambda-1}\ \omega_{\Lambda-2}}\ .$$
To have the same type of the indices at $(\Lambda)$ and
$(\Lambda-1)$ terms, we also change the summation index $j$ to $\mu-j$
in Eq. (\ref{mtc2}). After these changes we read:
\begin {equation}
\left(\Lambda\right)^{2\mu}_{p}= \sum\limits_{j=0}^{\mu}\;
\binom{\mu}{j}\left(\frac{1}{\omega_{\Lambda-1}}\right)^{2\mu-2j}
\sum\limits_{q=\max{(0,\ p-2\mu+2j})}^{2j}\ a_{p}^{2\mu-2j+q}\
(\Lambda-1)^{2j}_q \left(\frac{\sigma^2}{\omega_{\Lambda-1}\
\omega_{\Lambda-2}}\right)^q \label{mtc3}
\end {equation}
We would like to represent the recurrence relation (\ref{mtc3}) as a
product of the matrices, therefore we introduce the new summation
indices $\lambda$ and $i$ by prescriptions:
$$\lambda = j,\ i = 2\lambda-q\ ,$$
and by change of the order of the summations and by the index $p$
transformation $$p\rightarrow 2\mu-p$$ we find:
\begin {equation}
\left(\Lambda\right)^{2\mu}_{2\mu-p}= \sum\limits_{i=0}^{p}\
\sum\limits^{\mu}_{\lambda = [\frac{i+1}{2}]}
\binom{\mu}{\lambda}\left(\frac{1}{\omega_{\Lambda-1}}\right)^{2\mu-2\lambda}
\ a_{2\mu-p}^{2\mu-i}\ (\Lambda-1)^{2\lambda}_{2\lambda-i}
\left(\frac{\sigma^2}{\omega_{\Lambda-1}\
\omega_{\Lambda-2}}\right)^{2\lambda-i} \label{mtc4}
\end {equation}

Then, in the spirit of the Appendix D, where we use the identity:
$$\omega_n = \frac{q_{n+1}}{q_n} = \frac{\sigma Q_{n+1}}{Q_n}\ ,$$
and we simplify the relation:
$$\left(\frac{1}{\omega_{\Lambda-1}}\right)^{2\mu-2\lambda}\ \left(\frac{\sigma^2}{\omega_{\Lambda-1}\
\omega_{\Lambda-2}}\right)^{2\lambda-i} =
Q^{4\mu-4\lambda}_{\Lambda-1}\ \frac{\left(\sigma Q_{\Lambda-2}\
Q_{\Lambda-1} \right)^{2\lambda-i}}{\left(\sigma Q_{\Lambda-1}\
Q_{\Lambda} \right)^{p-i}\ \left(\sigma Q_{\Lambda-1}\ Q_{\Lambda}
\right)^{2\mu-p}}\ .$$
Inserting this identity to Eq. (\ref{mtc4}) we
find for the equation (\ref{mtc2}) the following result:
\begin {eqnarray}
& &\left(\sigma Q_{\Lambda-1}\ Q_{\Lambda} \right)^{2\mu-p}\
\left(\Lambda\right)^{2\mu}_{2\mu-p}=\label{mtc5} \\
&=&\sum\limits_{i=0}^{p}\ \sum\limits^{\mu}_{\lambda =
[\frac{i+1}{2}]} \frac{a_{2\mu-p}^{2\mu-i}}{\left(\sigma
Q_{\Lambda-1}\ Q_{\Lambda} \right)^{p-i}}\ \left(\sigma
Q_{\Lambda-2}\ Q_{\Lambda-1}
\right)^{2\lambda-i}(\Lambda-1)^{2\lambda}_{2\lambda-i}\
\binom{\mu}{\lambda}Q^{4\mu-4\lambda}_{\Lambda-1}\nonumber
\end {eqnarray}

On the left-hand side of the above equation we have the matrix element
of the $\mu-th$ column of the matrix defined as the product of three matrices.
Let us define an auxiliary matrix $\ms X^d(\Lambda)$ by the equation:

\begin{equation}
\ms X^d_{p,\rho}(\Lambda) = \sum\limits_{i=0}^p
\sum\limits^{\rho}_{\lambda=[\frac{i+1}{2}]}\; \ms
A^d_{p,i}(\Lambda-1)\ms C^d_{i,\lambda}(\Lambda-1)\ms
M^d_{\lambda,\rho}(\Lambda-1) ,\label{mtc51}
\end{equation}
The index $d$ correspond to the dimension of the principal nonzero minor of the
each of the matrices.

 Let us suppose, that matrix $\ms C^d(\Lambda-1)$ is known and we are going to evaluate the matrix $\ms C^d(\Lambda)$
 following the Eq. (\ref{mtc51}). Following the recurrence relation (\ref{mtc5}), the matrix $\ms C^{\mu}(\Lambda)$ possesses
 from $\ms X^d(\Lambda)$ $d-th$ column only. This means, that to obtain
 the whole matrix $\ms C^{\mu}(\Lambda)$ we must to evaluate matrices $\ms X^d_{p,d}(\Lambda)$ for all $d=0, 1, \cdots , \mu,$
 for each such matrix to strip of the $d-th$ column. Such column will be $d-th$ column of the matrix $\ms C^{\mu}(\Lambda).$
  This linear operation is expressed as:
$$\ms C^{\mu}(\Lambda) = \sum\limits^{\mu}_{d=0}\ \ms X^{d}(\Lambda)\ \ms P^d\ ,$$
where $\ms P^d$ is the projector of the $d-th$ column:
$$\left\{\ms P^d\right\}_{\lambda,q} = \delta_{d,\lambda}\ \delta_{\lambda,q}\ .$$

We define the matrices as follows.

\noindent $\ms A^d$ and $\ms M^d$ are the matrices of the dimensions
$(2\mu+1)(2\mu+1)$ and $(\mu+1)(\mu+1)$ respectively,  $\ms C^d$ is
the matrix of dimensions $(2\mu+1)(\mu+1)$. These matrices possess
nonzero main minors of the dimensions $(2d+1)(2d+1)$, $(d+1)(d+1)$,
and $(2d+1)(d+1),$ respectively.

1. The matrix $\ms A^d(\Lambda-1)$ of the dimension
$(2\mu+1)(2\mu+1)$ is defined as
$$\left\{\ms A^d(\Lambda-1)\right\}_{p,i} = \frac{a_{2d-p}^{2d-i}}{\left(\sigma
Q_{\Lambda-1}\ Q_{\Lambda} \right)^{p-i}}\ .$$ From the definition
of the symbol $a_{2\mu-p}^{2\mu-i}$ in Eq. (\ref{recur42}) we see,
that $\ms A^d(\Lambda-1)$ is the lower-triangular matrix with
non-zero main minor of the dimension $(2d+1)(2d+1)\ .$ The
lower-triangularity of this matrix results in the upper bound of the
summation over index $i$ in Eq. (\ref{mtc51}).

2. The matrix $\ms M^d(\Lambda-1)$ of the dimension $(\mu+1)(\mu+1)$
is defined as
$$\left\{\ms M^d(\Lambda-1)\right\}_{\lambda,q} = \binom{q}{\lambda}Q^{4q-4\lambda}_{\Lambda-1}\Theta(d,q)\ ,$$
where $$\Theta(d,q) = 1, \ q \leq d;\ \Theta(d,q) = 0, \ q > d.$$
From the definition of the binomial factor $\binom{q}{\lambda}$ we
see, that $\ms M^d(\Lambda-1)$ is the upper-triangular matrix of the
dimension $(\mu+1)(\mu+1)$ with non-zero main minor of the dimension
$(d+1)(d+1)\ .$

3. The characteristics of the matrix $\ms C^d(\Lambda-1)$ can be
deduced from the product of two matrices:
$$\left\{\ms C^d(\Lambda-1)\ \ms M^d(\Lambda-1)\right\}_{i,q} =
\sum\limits^{\mu}_{\lambda=[\frac{i+1}{2}]}\left\{\ms
C^d(\Lambda-1)\right\}_{i,\lambda} \left\{\ms
M^d(\Lambda-1)\right\}_{\lambda,q}\ .$$ Because index $\lambda$ in
the above relation runs from $[\frac{i+1}{2}]$ we define that matrix
$\ms C^d(\Lambda-1)$  in the $i-th$ row possesses the zeros up to
$[\frac{i+1}{2}]-th$ term. This matrix is the upper-triangular of
the dimension $(2\mu+1)(\mu+1)$ with the nonzero main minor of the
dimension $(2d+1)(d+1)\ .$ The connection of the matrix element of
the $\left\{\ms C^{\mu}(\Lambda)\right\}_{p,\mu}$ with the quantity
$\left(\Lambda\right)^{2\mu}_{2\mu-p}$ in the relation (\ref{mtc1})
for evaluation of the $N-$ dimensional integral is:
\begin {equation}
\left\{\ms C^{\mu}(\Lambda)\right\}_{p,\mu} = \left(\sigma\
Q_{\Lambda}Q_{\Lambda-1}\right)^{2\mu-p}\left(\Lambda\right)^{2\mu}_{2\mu-p}
\label{mtc9}
\end {equation}

Finally, we have for the matrix $\ms C^{\mu}(\Lambda)$ the relation:
$$\ms C^{\mu}(\Lambda) = \sum\limits^{\mu}_{i_{\Lambda}=0}\ \ms A^{i_{\Lambda}}(\Lambda-1)\
\ms C^{i_{\Lambda}}(\Lambda-1)\ \ms M^{i_{\Lambda}}(\Lambda-1)\ \ms
P^{i_{\Lambda}}\ .$$

After the full recurrence procedure we have for the value $\ms
C^{\mu}(\Lambda)$ the relation:
\begin {eqnarray}
& &\ms C^{\mu}(\Lambda) = \sum\limits^{\mu}_{i_{\Lambda}=0}\
\sum\limits^{i_{\Lambda}}_{i_{\Lambda-1}=0}\ \cdots\
\sum\limits^{i_3}_{i_{2}=0} \ms A^{i_{\Lambda}}(\Lambda-1)\ \ms
A^{i_{\Lambda-1}}(\Lambda-2)\ \ms A^{i_{\Lambda-2}}(\Lambda-3)\
\cdots\ \ms A^{i_2}(1)\label{mtc10} \\
& &\ms C^{i_2}(1)\ \tilde{\ms M}^{i_2}(1)\ \tilde{\ms M}^{i_3}(2)\
\cdots\ \tilde{\ms M}^{i_{\Lambda-1}}(\Lambda-2)\ \tilde{\ms
M}^{i_{\Lambda}}(\Lambda-1) \nonumber
\end {eqnarray}
where $\ms C^{i_2}(1)$ is defined as (\ref{mtc9}, \ref{mtc2}):
$$\left\{\ms C^{i_2}(1)\right\}_{p,i_2} = (\sigma\ Q_1Q_0)^{2i_2-p}\ a^{2i_2}_{2i_2-p}\ .$$
The value $\tilde{\ms M}^{i}(K)$ is the matrix with nonzero $i-th$
column defined as:
$$\tilde{\ms M}^{i}(K) = \ms M^{i}(K)\ \ms P^i\ .$$

To evaluate the Eq. (\ref{mtc10}) we profit from the associative law of the
products of the matrices.

To evaluate the product
$$\left\{\ms A^{i_{\Lambda}}(\Lambda-1)\ \ms
A^{i_{\Lambda-1}}(\Lambda-2)\ \ms A^{i_{\Lambda-2}}(\Lambda-3)\
\cdots\ \ms A^{i_2}(1)\right\}_{p,\lambda}$$ we use the identity:
$$\left\{\ms A^{i_3}(2)\ \ms A^{i_2}(1)\right\}_{p,\lambda}\
=\ 2^{2\lambda-2p}\frac{(4i_2-2\lambda)!}{(4i_3-2p)!(p-\lambda)!}\
\lim_{x\rightarrow 1}\ \partial_x^{4i_3-4i_2}\
\left[x^{4i_3-2p}\left(\frac{x^2}{\sigma Q_3Q_2}+\frac{1}{\sigma
Q_2Q_1}\right)^{p-\lambda}\right]$$

Proof of the identity:

\noindent The product of the above lower-diagonal matrices is:
$$\left\{\ms A^{i_3}(2)\ \ms A^{i_2}(1)\right\}_{p,\lambda}\
=\ \sum\limits_{j=\lambda}^p \left\{\ms A^{i_3}(2)\right\}_{p,j}
\left\{\ms A^{i_2}(1)\right\}_{j,\lambda} =
\sum\limits_{j=\lambda}^p\ \frac{a^{2i_3-j}_{2i_3-p}}{(\sigma\
Q_3Q_2)^{p-j}}\ \frac{a^{2i_2-\lambda}_{2i_2-j}}{(\sigma\
Q_2Q_1)^{j-\lambda}}\ .$$ Following the definition of the value
$$a^{2i_3-j}_{2i_3-p} = \binom{2i_3-j}{2i_3-p}\ \frac{\pochh{1/2}{2i_3-j}}{\pochh{1/2}{2i_3-p}}\ ,$$
after some algebra we find:
$$\left\{\ms A^{i_3}(2)\ \ms A^{i_2}(1)\right\}_{p,\lambda}\
=\ 2^{2\lambda-2p}\ \frac{(4i_2-2\lambda)!}{(4i_3-2p)!(p-\lambda)!}\
\sum\limits_{j=\lambda}^p\ \binom{p-\lambda}{j-\lambda}\
\frac{(4i_3-2j)!}{(4i_2-2j)!}\ \frac{1}{(\sigma\ Q_3Q_2)^{p-j}}\
\frac{1}{(\sigma\ Q_2Q_1)^{j-\lambda}}\ .
$$
We use the identity, where the derivative is not dependent on the
summation index $j$:
$$\frac{(4i_3-2j)!}{(4i_2-2j)!}\ =\ \lim_{x\rightarrow 1}\ \partial_x^{4i_3-4i_2}\
\left(x^{4i_3-2j}\right)\ ,$$ we obtain for the product of the
matrices the relation:
\begin {eqnarray}
& &\left\{\ms A^{i_3}(2)\ \ms A^{i_2}(1)\right\}_{p,\lambda}\ =\nonumber\\
&=& 2^{2\lambda-2p}\frac{(4i_2-2\lambda)!}{(4i_3-2p)!(p-\lambda)!}\
\lim_{x\rightarrow 1}\ \partial_x^{4i_3-4i_2}\ \left[x^{4i_3-2p}\
\sum\limits_{j=\lambda}^p\ \binom{p-\lambda}{j-\lambda}\
\left(\frac{x^2}{\sigma Q_3Q_2}\right)^{p-j}\ \left(\frac{1}{\sigma
Q_2Q_1}\right)^{j-\lambda}\right]\ ,\nonumber
\end {eqnarray}
after the summation over index $j$ we obtain the above identity.

\noindent QED.

For product of the tree upper diagonal matrices $\left\{\ms
A^{i_4}(3)\ \ms A^{i_3}(2)\ \ms A^{i_2}(1)\right\}$ by the same
evaluation methods we find:
\begin {eqnarray}
& &\left\{\ms A^{i_4}(3)\ \ms A^{i_3}(2)\ \ms
A^{i_2}(1)\right\}_{k,\lambda}\ = \sum^k_{p=\lambda}\ \left\{\ms
A^{i_4}(3)\right\}_{k,p}\left\{ \ms A^{i_3}(2)\ \ms
A^{i_2}(1)\right\}_{p,\lambda}\ =
2^{2\lambda-2k}\frac{(4i_2-2\lambda)!}{(4i_4-2k)!(k-\lambda)!}\\
& &\lim_{x\rightarrow 1}\ \partial_x^{4i_3-4i_2}\
\left\{x^{4i_3-2k}\ \lim_{y\rightarrow 1}\ \partial_y^{4i_4-4i_3}\
\left\{y^{4i_4-2k}\left(\frac{x^2y^2}{\sigma\ Q_4Q_3} +
\frac{x^2}{\sigma\ Q_3Q_2} + \frac{1}{\sigma\ Q_2Q_1}
\right)^{k-\lambda}\right\}\right\}\nonumber
\end {eqnarray}

By mathematical induction we can prove for product of $n$
lower-triangular matrices $\ms A^i$ of the dimension
$(2\mu+1)(2\mu+1)$ with non-zero main minor of the dimension
$(2i+1)(2i+1)$:

\textit{Lemma:} The matrix elements of the product of $n$ matrices
$$\ms A^{i_{n+1}}(n)\ \ms
A^{i_{n}}(n-1)\ \ms A^{i_{n-1}}(n-2)\ \cdots\ \ms A^{i_3}(2) \ms
A^{i_2}(1)$$ are given by relation:
\begin {eqnarray}
& &\left\{\ms A^{i_{n+1}}(n)\ \ms A^{i_{n}}(n-1)\ \ms A^{i_{n-1}}(n-2)\
\cdots\ \ms A^{i_3}(2) \ms
A^{i_2}(1)\right\}_{k,\lambda} =\\
& &2^{2\lambda-2k}\frac{(4i_2-2\lambda)!}{(4i_{n+1}-2k)!(k-\lambda)!}
\lim_{x_1, x_2, \cdots , x_{n-1}\rightarrow 1}\
\left(\partial^4_{x_1}\right)^{i_3-i_2}\
\left(\partial^4_{x_2}\right)^{i_4-i_3}\ \cdots\
\left(\partial^4_{x_{n-1}}\right)^{i_{n+1}-i_n}\nonumber \\
& &\left\{x_1^{4i_3-2k}\ x_2^{4i_4-2k}\ \cdots\ x_{n-1}^{4i_{n+1}-2k}\
\left(\frac{x_1^2 x_2^2 \cdots x^2_{n-2} x^2_{n-1}}{\sigma
Q_{n+1}Q_{n}}\ +\ \frac{x_1^2 x_2^2 \cdots
 x^2_{n-2}}{\sigma Q_{n}Q_{n-1}}\ +\ \cdots\ +\ \frac{1}{\sigma Q_{2}Q_{1}} \right)^{k-\lambda}\right\}\ .\nonumber
\end {eqnarray}

We are going evaluate the product of the matrices $\tilde{\ms
M}^{d}(K)\ .$ The matrix element is evaluated as:
$$\left\{\tilde{\ms
M}^{d}(K)\right\}_{\lambda,q} = \sum\limits^{\mu}_{i=0}\ \left\{\ms
M^d(K)\right\}_{\lambda,i}\left\{\ms P^d\right\}_{i,q}\ =\
\sum\limits^{\mu}_{i=0}\binom{i}{\lambda}\ Q^{4i-4\lambda}_K\
\delta_{d,i}\ \delta_{d,q}\ =\ \binom{d}{\lambda}\
Q^{4d-4\lambda}_K\ \delta_{d,q}\ .$$ Matrix $\tilde{\ms M}^{d}(K)$
is the $(\mu+1)(\mu+1)$ dimensional, upper-triangular, with $d-th$
nonzero column. Following this identity, we find:
$$\left\{\tilde{\ms M}^{i_2}(1)\ \tilde{\ms M}^{i_3}(2)\right\}_{k,\lambda} =
\sum\limits^{\mu}_{i=0}\ \left\{\tilde{\ms
M}^{i_2}(1)\right\}_{k,i}\left\{ \tilde{\ms
M}^{i_3}(2)\right\}_{i,\lambda}\ =\ \binom{i_2}{k}\binom{i_3}{i_2}\
Q_1^{4i_2-4k}Q_2^{4i_3-4i_2}\ \delta_{i_3,\lambda}\ .$$ For product
of tree matrices we find:
\begin {eqnarray}
&&\left\{\tilde{\ms M}^{i_2}(1)\ \tilde{\ms M}^{i_3}(2)\ \tilde{\ms
M}^{i_4}(3)\right\}_{k,p} =
\sum\limits^{\mu}_{\lambda=0}\left\{\tilde{\ms M}^{i_2}(1)\
\tilde{\ms M}^{i_3}(2)\right\}_{k,\lambda}\left\{\tilde{\ms
M}^{i_4}(3)\right\}_{\lambda,p}\ =\nonumber \\
&&\binom{i_2}{k}\binom{i_3}{i_2}\binom{i_4}{i_3}\
Q_1^{4i_2-4k}Q_2^{4i_3-4i_2}Q_3^{4i_4-4i_3}\ \delta_{i_4,p}\ .
\end {eqnarray}

By mathematical induction we can prove for product of $n$ upper-triangular matrices
$\tilde{\ms M}^{i}$ of the dimension
$(\mu+1)(\mu+1)$ with non-zero main minor of the dimension
$(i+1)(i+1)$:

\textit{Lemma:} The matrix elements of the product of $n$ matrices
$$\tilde{\ms M}^{i_{2}}(1)\ \tilde{\ms M}^{i_{3}}(2)\ \tilde{\ms M}^{i_{4}}(3)\
\cdots\ \tilde{\ms M}^{i_n}(n-1) \tilde{\ms M}^{i_{n+1}}(n)$$ are given by relation:

\begin {eqnarray}
&&\left\{\tilde{\ms M}^{i_{2}}(1)\ \tilde{\ms M}^{i_{3}}(2)\ \tilde{\ms M}^{i_{4}}(3)\
\cdots\ \tilde{\ms M}^{i_n}(n-1) \tilde{\ms M}^{i_{n+1}}(n)\right\}_{\lambda,p}\ = \\
&&\binom{i_2}{\lambda}\binom{i_3}{i_2}\binom{i_4}{i_3}\ \cdots\ \binom{i_{n+1}}{i_n}
Q_1^{4i_2-4\lambda}Q_2^{4i_3-4i_2}Q_3^{4i_4-4i_3}\ \cdots\ Q_n^{4i_{n+1}-4i_n}\ \delta_{i_{n+1},p}\ .\nonumber
\end {eqnarray}

Now, inserting all this into Eq. (\ref{mtc10}), we find for the matrix elements of the matrix $\ms C^{\mu}(\Lambda)$ the relation:

\begin {eqnarray}
& &\left\{\ms C^{\mu}(\Lambda)\right\}_{k,p}\ =\ \sum\limits^{\mu}_{i_{\Lambda}=0}\
\sum\limits^{i_{\Lambda}}_{i_{\Lambda-1}=0}\ \cdots\ \sum\limits^{i_3}_{i_{2}=0}
\sum\limits^{2\mu}_{\lambda=0} \sum\limits^{\mu}_{z=[\frac{\lambda+1}{2}]}\\
& &\left\{\ms A^{i_{\Lambda}}(\Lambda-1)\ \ms
A^{i_{\Lambda-1}}(\Lambda-2)\ \cdots\ \ms A^{i_2}(1)\right\}_{k,\lambda}
\left\{\ms C^{i_2}(1)\right\}_{\lambda,z}
\left\{\tilde{\ms M}^{i_2}(1)\
\cdots\ \tilde{\ms M}^{i_{\Lambda-1}}(\Lambda-2)\ \tilde{\ms
M}^{i_{\Lambda}}(\Lambda-1)\right\}_{z,p}\ = \nonumber \\
&=&\sum\limits^{\mu}_{i_{\Lambda}=0}\
 \cdots\ \sum\limits^{i_3}_{i_{2}=0}
\sum\limits^{2\mu}_{\lambda=0} \sum\limits^{\mu}_{z=[\frac{\lambda+1}{2}]}2^{2\lambda-2k}\frac{(4i_2-2\lambda)!}{(4i_{\Lambda}-2k)!(k-\lambda)!}
\lim_{x_1, \cdots , x_{\Lambda-2}\rightarrow 1}\
\left(\partial^4_{x_1}\right)^{i_3-i_2}\
 \cdots\
\left(\partial^4_{x_{\Lambda-2}}\right)^{i_{\Lambda}-i_{\Lambda-1}}\nonumber \\
& &\left\{x_1^{4i_3-2k}\  \cdots\ x_{\Lambda-2}^{4i_{\Lambda}-2k}\
\left(\frac{x_1^2 \cdots x^2_{\Lambda-3} x^2_{\Lambda-2}}{\sigma
Q_{\Lambda}Q_{\Lambda-1}}\ +\ \frac{x_1^2  \cdots
 x^2_{\Lambda-3}}{\sigma Q_{\Lambda-1}Q_{\Lambda-2}}\ +\ \cdots\ +\ \frac{1}{\sigma Q_{2}Q_{1}} \right)^{k-\lambda}\right\}\
\left\{\ms C^{i_2}(1)\right\}_{\lambda,z} \nonumber \\
& &\binom{i_2}{z}\binom{i_3}{i_2}\binom{i_4}{i_3}\ \cdots\ \binom{i_{\Lambda}}{i_{\Lambda-1}}
Q_1^{4i_2-4z}Q_2^{4i_3-4i_2}Q_3^{4i_4-4i_3}\ \cdots\ Q_{\Lambda-1}^{4i_{\Lambda}-4i_{\Lambda-1}}\ \delta_{i_{\Lambda},p}\ .\nonumber
\end {eqnarray}

In this relation, the nonzero minor of the $(2\mu+1)(\mu+1)$ dimensional matrix $\left\{\ms C^{i_2}(1)\right\}_{\lambda,z}$ is
$(2i_2+1)(i_2+1)$ dimensional and we can change the limits of the summations over indices $\lambda,z\ .$ Taking into account the $\delta_{i_{\Lambda},p}$ and the definitions (\ref{mtc9}), (\ref{mtc2}):

$$\left\{\ms C^{i_2}(1)\right\}_{\lambda,z}\ =\ (\sigma Q_1 Q_0)^{2z-\lambda}\ \left(1\right)^{2z}_{2z-\lambda}\ =\
(\sigma Q_1 Q_0)^{2z-\lambda}\ (a)^{2z}_{2z-\lambda}\ ,$$
we have the the common relation for the matrix element $\left\{\ms C^{\mu}(\Lambda)\right\}_{k,p}\ : $
\begin {eqnarray}
& &\left\{\ms C^{\mu}(\Lambda)\right\}_{k,p}\ =\
\sum\limits^{p}_{i_{\Lambda-1}=0}\ \cdots\ \sum\limits^{i_3}_{i_{2}=0}
\sum\limits^{2i_2}_{\lambda=0} \sum\limits^{i_2}_{z=[\frac{\lambda+1}{2}]}\label{mtc100} \\
& &2^{2\lambda-2k}\frac{(4i_2-2\lambda)!}{(4p-2k)!(k-\lambda)!}
\lim_{x_1, \cdots , x_{\Lambda-2}\rightarrow 1}\
\left(\partial^4_{x_1}\right)^{i_3-i_2}\
 \cdots\
\left(\partial^4_{x_{\Lambda-2}}\right)^{p-i_{\Lambda-1}}\nonumber \\
& &\left\{x_1^{4i_3-2k}\  \cdots\ x_{\Lambda-2}^{4p-2k}\
\left(\frac{x_1^2 \cdots x^2_{\Lambda-3} x^2_{\Lambda-2}}{\sigma
Q_{\Lambda}Q_{\Lambda-1}}\ +\ \frac{x_1^2  \cdots
 x^2_{\Lambda-3}}{\sigma Q_{\Lambda-1}Q_{\Lambda-2}}\ +\ \cdots\ +\ \frac{1}{\sigma Q_{2}Q_{1}} \right)^{k-\lambda}\right\}\ \nonumber \\
& &(\sigma Q_1 Q_0)^{2z-\lambda}\ (a)^{2z}_{2z-\lambda} \nonumber \\
& &\binom{i_2}{z}\binom{i_3}{i_2}\binom{i_4}{i_3}\ \cdots\ \binom{p}{i_{\Lambda-1}}
Q_1^{4i_2-4z}Q_2^{4i_3-4i_2}Q_3^{4i_4-4i_3}\ \cdots\ Q_{\Lambda-1}^{4p-4i_{\Lambda-1}}\ .\nonumber
\end {eqnarray}
This is desired relation for evaluation the the an-harmonic correction to the Moeler's formula for
the propagator of the harmonic oscillator.

\section{Evaluation of the continuum limit of the an-harmonic correction}

In the Appendix E we evaluated the quality $\left\{\ms
C^{\mu}(\Lambda)\right\}_{p,\mu}$, which is related to symbol
$\left(\Lambda\right)^{2\nu}_{2\nu-p}$  in Eq. (\ref{ch307}) for
$\Lambda = N-1$
 by the relation (\ref{mtc9}):
\begin {equation}
\left\{\ms C^{\mu}(\Lambda)\right\}_{p,\mu} = \left(\sigma\
Q_{\Lambda}Q_{\Lambda-1}\right)^{2\mu-p}\left(\Lambda\right)^{2\mu}_{2\mu-p}\
. \label{ch308}
\end {equation}

Finally, we are going to evaluate the continuum limit of the
expression:
\begin {equation}
 \triangle^{3\nu}\sum \limits_{p=0}^{2\nu}\ \frac{(\xi)^{2\nu-p}}{\left(\sigma\
Q_{N-1}Q_{N-2}\right)^{2\nu-p}}\ \left\{\ms
C^{\nu}(N-1)\right\}_{p,\nu}\ . \label{ch309}
\end {equation}

To obtain in the continuum limit the non-vanishing finite expression
we must to show that in the above sum are the terms, which are
divergent as $\triangle^{-3\nu} .$ All terms with  step of
divergence less than $\triangle^{-3\nu}$ will be wiped out in the
continuum limit by the factor $\triangle^{3\nu}$ in front of the
sum. Moreover we show that there are not the term with stronger
divergence as $\triangle^{-3\nu} .$

Due to the definition in (\ref{recur101}):
$$\xi = \frac{1}{\omega_{N-2}}\
\frac{c}{4\triangle(1+\frac{b\triangle^2}{c})}\ \varphi_N^2$$ we
have a contribution to the power of the divergence
$$\xi^{2\nu-p}\ \sim \ \frac{1}{\triangle^{2\nu-p}}\ .$$

In Eq. (\ref{mtc100}), we can find the following powers of divergent term $1/\triangle$ :

The $Q_n$ and $\tilde{Q}_n$ are introduced in Appendix D by the
relations (\ref{expo22}, \ref{expo23}):
$$Q_n = \left(\frac{\rho_1}{\sigma}\right)^{n} + \frac{\tilde{u}_2}{\tilde{u}_1} \left(\frac{\rho_2}{\sigma}\right)^{n}\ ,$$
$$\tilde{Q}_n = \left(\frac{\rho_1}{\sigma}\right)^{n} - \frac{\tilde{u}_2}{\tilde{u}_1} \left(\frac{\rho_2}{\sigma}\right)^{n}\ .$$

Taking into account the identity:
\begin {equation}
\frac{\tilde{Q}_n}{Q_n}-\frac{\tilde{Q}_{n-1}}{Q_{n-1}} =
\frac{2\frac{\tilde{u}_2}{\tilde{u}_1}(\rho_1-\rho_2)}{\sigma Q_n
Q_{n-1}}\ ,\label{ch310a}
\end {equation}

we see that
$$\frac{1}{\sigma Q_n Q_{n-1}}=\frac{\tilde{u}_1}{2\tilde{u}_2(\rho_1-\rho_2)}
\left(\frac{\tilde{Q}_n}{Q_n}-\frac{\tilde{Q}_{n-1}}{Q_{n-1}}\right)$$
where, following Eq. (\ref{expo1})and (\ref{expo2}):
$$\frac{\tilde{u}_1}{2\tilde{u}_2(\rho_1-\rho_2)}\ =\
-\frac{1}{2\triangle \gamma}\ (1+ \mathcal{O}(\triangle))\ .$$

We can conclude, that the expression
$$\lim_{x_1, \cdots , x_{\Lambda-2}\rightarrow 1}\
\left(\partial^4_{x_1}\right)^{i_3-i_2}\
 \cdots\
\left(\partial^4_{x_{\Lambda-2}}\right)^{\nu-i_{\Lambda-1}}
\left\{x_1^{4i_3-2p}\  \cdots\ x_{\Lambda-2}^{4\nu-2p}\
\left(\frac{x_1^2 \cdots x^2_{\Lambda-3} x^2_{\Lambda-2}}{\sigma
Q_{\Lambda}Q_{\Lambda-1}}\  +\ \cdots\ +\ \frac{1}{\sigma
Q_{2}Q_{1}} \right)^{p-\lambda}\right\}$$ in the continuum limit
diverges as
$$\frac{1}{\triangle^{p-\lambda}}\ .$$

The third source of the powers of the factors $1/\triangle$ we find
from the re-ordering of the summations
$$\sum\limits^{\nu}_{i_{\Lambda-1}=0}\ \cdots\
\sum\limits^{i_3}_{i_{2}=0}\ .$$
We arrange the set of the summation
indices into the groups, where the indices are equal and let indices
$\{i_{j_1}, i_{j_2},\cdots\ , i_{j_{\mu}}\}$ are the first indices
of the new equal valued group in the descendant order:
$\left\{i_{\Lambda-1}, i_{\Lambda-2}, \cdots , i_{j_1+1}\right\},$
$\left\{i_{j_1}, i_{j_1-1}, \cdots , i_{j_2+1}\right\},$ $\cdots,$
$\left\{i_{j_{\mu}}, i_{j_{\mu}-1}, \cdots , i_{2}\right\}\ .$
We divided the all indices to $\mu+1$ groups.
 For the equal indices from the same group $i_j = i_{j-1}$ in
Eq.(\ref{mtc100}) disappears the derivatives
$$\left(\partial^4_{x_{(j-2)}}\right)^{(i_j - i_{j-1})}$$ and the
powers of $$Q_{j-1}^{4(i_j - i_{j-1})}\ .$$
When the derivative over some variable disappears, there is possible to provide limit
over such variable, and this variable don't act in the expression.

For every set of the
numbers $\left\{j_1, j_2, \cdots , j_{\mu}\right\}$ we have
contributions to Eq. (\ref{mtc100}) in the form:

\begin {eqnarray}
& &\left\{\ms C^{\nu}(\Lambda)\right\}_{p,\nu}(j_1, j_2, \cdots ,
j_{\mu})\ =\ 2^{2\lambda-2p}\frac{(4i_{(j_{\mu})
}-2\lambda)!}{(4\nu-2p)!(p-\lambda)!}  \lim_{x_{j_{\mu}}, \cdots
, x_{j_2}, x_{j_1}\rightarrow 1}\
\left(\partial^4_{x_{j_{\mu}}}\right)^{(i_{j_{\mu-1}}-i_{j_{\mu}})}\
 \cdots\
 \left(\partial^4_{x_{j_1}}\right)^{(\nu-i_{j_1})} \nonumber \\
& &\left\{x_{j_{\mu}}^{4i_{(j_{\mu-1)}}-2p}\  \cdots \
x_{j_1}^{4\nu-2p}\ \left((x_{j_{\nu}}^2 \cdots x^2_{j_2}
x^2_{j_1})\left(\frac{1}{\sigma Q_{\Lambda}Q_{\Lambda-1}}+
\frac{1}{\sigma Q_{\Lambda-1}Q_{\Lambda-2}}+\ \cdots\ +
\frac{1}{\sigma
Q_{j_1+1}Q_{j_1}}\right)\ +\ \right. \right.\label{ch329} \\
& & \left. \left. (x_{j_{\nu}}^2 \cdots x^2_{j_2})
\left(\frac{1}{\sigma Q_{j_1}Q_{j_1-1}}+\ \cdots\ + \frac{1}{\sigma
Q_{j_2+1}Q_{j_2}}\right)\ +\ \cdots\ +\frac{1}{\sigma
Q_{j_{\nu}}Q_{j_{\nu}-1}}\ +\ \cdots\ +\
 \frac{1}{\sigma Q_{2}Q_{1}} \right)^{p-\lambda}\right\}\ \nonumber \\
& &(\sigma Q_1 Q_0)^{2z-\lambda}\ (a)^{2z}_{2z-\lambda}
\binom{i_{j_{\mu}}}{z}\binom{i_{j_{\mu-1}}}{{i_{j_{\mu}}}}\binom{i_{j_{\mu-2}}}{{i_{j_{\mu-1}}}}\
\cdots\ \binom{\nu}{i_{j_{1}}}
Q_{j_{\mu}}^{4i_{j_{\mu}}-4z}Q_{j_{(\mu-1)}}^{4i_{j_{(\mu-1)}}-4i_{j_{\mu}}}\
\cdots\ Q_{j_1}^{4\nu-4i_{\Lambda-1}}\ .\nonumber
\end {eqnarray}
To obtain the value (\ref{mtc100}), we must sum over all allowed indexes $j_i.$

In the continuum limit the summations are converted to integral by
the following prescription:
$$\sum\limits_{k=1}^N\ f(k)\ =\ \frac{1}{\triangle}\ \sum\limits_{k=1}^N\ \triangle f(k)\ \rightarrow
\frac{1}{\triangle}\ \int_0^{N \triangle}\ d\tau\ f(\tau)\ ,$$ where
$\tau = k \triangle.$

The  power of the divergent term $\frac{1}{\triangle}$  is equal to the number of integrations  $\mu,$ but this number is smaller or equal to $\nu.$
Altogether, the power of the factor $\frac{1}{\triangle}$ in the sum
of Eq. (\ref{ch309}) is equal to $(2\nu+\mu-\lambda).$ therefore,
we will have the finite nonzero contribution to the continuum limit
only for $\lambda\ =\ 0,$ and $\mu=\nu$. Practically, the
indices $i_j$ in the sum (\ref{mtc100}) are divided into $\nu+1$
groups, with equal values inside each group and the difference one
for neighbouring  groups. This means, that indices in the first group are equal to $\nu,$ and in the last group
are equal to zero. As the result than $i_{j_{\mu}}=i_2=0$ and the indexes $z=0$ also.

Then, the leading term of Eq. (\ref{mtc100}) in the continuum
limit is the sum of the contributions of Eq. (\ref{ch329}) over all
possible combinations of the indices $(j_1, j_2, \cdots , j_{\nu}).$
The leading term can be reads:
\begin {eqnarray}
& &\left\{\ms C^{\nu}(\Lambda)\right\}_{p,\nu}\ =\
2^{-2p}\frac{\nu!}{(4\nu-2p)!(p)!}
\sum\limits_{j_1=1}^{\Lambda-\nu}\
\sum\limits_{j_2=j_1+1}^{\Lambda-\nu+1}\ \cdots\
\sum\limits_{j_{\nu}=j_{(\nu-1)}+1}^{\Lambda}\ \lim_{x_{j_{\nu}},
\cdots , x_{j_2}, x_{j_1}\rightarrow 1}\
\left(\partial^4_{x_{j_{\nu}}}\right)\
 \cdots\
\left(\partial^4_{x_{j_2}}\right)\ \left(\partial^4_{x_{j_1}}\right) \label{ch330} \\
& &\left\{x_{j_{\nu}}^{4-2p}\ \cdots\ x_{j_{i}}^{4(\nu-i+1)-2p}\  \cdots\
x_{j_2}^{4(\nu-1)-2p}\ x_{j_1}^{4\nu-2p}\ \left((x_{j_{\nu}}^2
\cdots x^2_{j_2} x^2_{j_1})\left(\frac{1}{\sigma
Q_{\Lambda}Q_{\Lambda-1}}+ \ \cdots\ + \frac{1}{\sigma
Q_{j_1+1}Q_{j_1}}\right)\ +\ \right. \right.\nonumber \\
& & \left. \left. (x_{j_{\nu}}^2 \cdots x^2_{j_2})
\left(\frac{1}{\sigma Q_{j_1}Q_{j_1-1}}+\ \cdots\ + \frac{1}{\sigma
Q_{j_2+1}Q_{j_2}}\right)\ +\ \cdots\ +\frac{1}{\sigma
Q_{j_{\nu}}Q_{j_{\nu}-1}}\ +\ \cdots\ +\
 \frac{1}{\sigma Q_{2}Q_{1}} \right)^{p}\right\}\ \nonumber \\
& &Q_{j_{\nu}}^4\ Q_{j_{\nu}-1}^4\  \cdots\ Q_{j_2}^4\ Q_{j_1}^4\
.\nonumber
\end {eqnarray}

To complete the evaluations, we  evaluate the
derivatives over the auxiliary variables $x_j$ in above relation
(\ref{ch330}). By substitution $x^2_j\ =\ y_j$ we introduce the
operator $\mathcal{D}_y$:
$$\mathcal{D}_{y_j}\ =\partial^4_{x_j}\ =\
2^4\left(\frac{3}{4}\partial^2_{y_j}\ +\ 3\partial^3_{y_j}\ +\
\partial^4_{y_j}\right).$$

\noindent This operator acts on the function $y_1^{2\nu-k}\ f^k(y_1)$ as:
$$\mathcal{D}_{y_1}\ \left(y_1^{2\nu-k}\ f^k(y_1)\right),\ f(y_1) = a_{j_1}+(y_{j_{\nu}}\ \cdots\
y_{j_{2}})d_{j_1}y_1,$$ where $a_{j_1},\ d_{j_1}$ are independent on variable $y_1$ and
defined by:

\begin {eqnarray}
& &a_{j_1}=(y_{j_{\nu}} \cdots y_{j_2})\left(\frac{1}{\sigma Q_{j_1}Q_{j_1-1}}+\ \cdots\ + \frac{1}{\sigma Q_{j_2+1}Q_{j_2}}\right)\ +\\
&+& \cdots\ +(y_{j_{\nu}} \cdots y_{j_3})\left(\frac{1}{\sigma Q_{j_2}Q_{j_2-1}}+\ \cdots\ + \frac{1}{\sigma Q_{j_3+1}Q_{j_3}}\right)\ +
\cdots +\left(\frac{1}{\sigma
Q_{j_{\nu}}Q_{j_{\nu}-1}}\ +\ \cdots\ +\
 \frac{1}{\sigma Q_{2}Q_{1}}\right)\nonumber
 \end {eqnarray}

$$ d_{j_1}=\ \left(\frac{1}{\sigma Q_{\Lambda}Q_{\Lambda-1}}+
\frac{1}{\sigma Q_{\Lambda-1}Q_{\Lambda-2}}+\ \cdots\ +
\frac{1}{\sigma Q_{j_1+1}Q_{j_1}}\right). $$

We use the identity:
\begin {eqnarray}
c_{\alpha}\partial_y^{\alpha}\ \left(y^{2\nu-k}\ f^k(y)\right) &=&
c_{\alpha}\sum\limits_{m=0}^{\alpha}\binom{\alpha}{m}\
\frac{\Gamma(k+1)}{\Gamma(k-m+1)}\ \frac{\Gamma(2\nu-k+1)}{\Gamma(2\nu-k-\alpha+m+1)}\
(y_{j_{\nu}}\ \cdots\ y_{j_{2}}d_{j_1})^m\ f^{k-m}(y)\
,\label{ch340}
\end {eqnarray}
where $\alpha=2, 3, 4,$ and $c_2=3/4, c_3=3, c_4=1\ .$

Applying this identity to derivative over variable $y_{j_1}$ in Eq.
(\ref{ch330}) we reads:
\begin {eqnarray}
& &2^{4\nu}\ \lim_{y_{j_{\nu}}, \cdots , y_{j_2}, y_{j_1}\rightarrow
1}\ \left(D_{y_{j_{\nu}}}\right)\
 \cdots\
\left(D_{y_{j_2}}\right)\ \left(D_{y_{j_1}}\right) \label{ch331} \\
& &\left\{y_{j_{\nu}}^{2-p}\ y_{j_{\nu-1}}^{2.2-p}  \cdots\
y_{j_2}^{2(\nu-1)-p}\ y_{j_1}^{2\nu-p}\ \left((y_{j_{\nu}} \cdots
y_{j_2} y_{j_1})\left(\frac{1}{\sigma Q_{\Lambda}Q_{\Lambda-1}}+
\frac{1}{\sigma Q_{\Lambda-1}Q_{\Lambda-2}}+ \cdots\ +
\frac{1}{\sigma
Q_{j_1+1}Q_{j_1}}\right) + \right. \right.\nonumber \\
& & \left. \left. (y_{j_{\nu}} \cdots y_{j_2}) \left(\frac{1}{\sigma
Q_{j_1}Q_{j_1-1}}+ \cdots + \frac{1}{\sigma
Q_{j_2+1}Q_{j_2}}\right) + \cdots +\left(\frac{1}{\sigma
Q_{j_{\nu}}Q_{j_{\nu}-1}} + \cdots +
 \frac{1}{\sigma Q_{2}Q_{1}}\right) \right)^{p}\right\}= \nonumber \\
&=&2^{4\nu} \sum_{\alpha_1=2}^4\
c_{\alpha_1}\sum\limits_{m_1=0}^{\alpha_1}\binom{\alpha_1}{m_1}\
\frac{\Gamma(p+1)}{\Gamma(p-m_1+1)}\ \frac{\Gamma(2\nu-p+1)}{\Gamma(2\nu-\alpha_1-(p-m_1)+1)}(d_{j_1})^{m_1}\ \nonumber \\
& &\lim_{y_{j_{\nu}}, \cdots ,y_{j_2} \rightarrow 1}\
\left(D_{y_{j_{\nu}}}\right)\
 \cdots\
\left(D_{y_{j_2}}\right)\nonumber   \\
& &\left\{y_{j_{\nu}}^{2-(p-m_1)}\ y_{j_{\nu-1}}^{2.2-(p-m_1)}
\cdots\ y_{j_2}^{2(\nu-1)-(p-m_1)} \left((y_{j_{\nu}} \cdots
y_{j_2}) \left(\frac{1}{\sigma Q_{\Lambda}Q_{\Lambda-1}}+
\frac{1}{\sigma Q_{\Lambda-1}Q_{\Lambda-2}}+\ \cdots\ +
\frac{1}{\sigma
Q_{j_2+1}Q_{j_2}}\right)+ \right. \right.\nonumber \\
&+& \left. \left. (y_{j_{\nu}} \cdots y_{j_3}) \left(\frac{1}{\sigma
Q_{j_2}Q_{j_2-1}}+ \cdots + \frac{1}{\sigma
Q_{j_3+1}Q_{j_3}}\right) + \cdots\ +\left(\frac{1}{\sigma
Q_{j_{\nu}}Q_{j_{\nu}-1}} + \cdots +
 \frac{1}{\sigma Q_{2}Q_{1}}\right) \right)^{(p-m_1)}\right\} \nonumber
\end {eqnarray}

After the limit $\lim{y_{j_1}}\rightarrow 1,$ we are left with
$f(1)= a_{j_1}+(y_{j_{\nu}} \cdots
y_{j_2})d_{j_1}\ ,\ $ where, according to the derivative with
the next variable $y_{j_{2}}$ the same procedure cam be applied with
the new constants $a_{j_{2}}, d_{j_{2}}.$ As the result of such
recurrence procedure we find:

\begin {eqnarray}
& &2^{4\nu}\ \sum_{\alpha_1=2}^4\ \cdots\ \sum_{\alpha_{\nu}=2}^4\
c_{\alpha_1}\cdots c_{\alpha_{\nu}}
\sum\limits_{m_1=0}^{\alpha_1}\
\sum\limits_{m_2=0}^{\alpha_2}\cdots
\cdots\sum\limits_{m_{\nu}=0}^{\alpha_{\nu}}
\binom{\alpha_1}{m_1}\binom{\alpha_2}{m_2}\cdots \binom{\alpha_{\nu}}{m_{\nu}}\ d^{m_1}_{j_1}\cdots\ d^{m_{\nu}}_{j_{\nu}}\label{ch332} \\
& &\frac{\Gamma(p+1)}{\Gamma(p-m_1-m_2-\cdots -m_{\nu}+1)}
\frac{\Gamma(2\nu-p+1)}{\Gamma(2-\alpha_{\nu}-(p-m_1-m_2-\cdots -m_{\nu})+1)}\frac{\Gamma(2(\nu-1)-(p-m_1)+1)}{\Gamma(2(\nu-1)-(p-m_1)-(\alpha_1-2)+1)} \nonumber \\
& &\cdots \frac{\Gamma(2-(p-m_1-m_2-\cdots
-m_{(\nu-1)})+1)}{\Gamma(2-(p-m_1-m_2-\cdots
-m_{(\nu-1)})-(\alpha_{\nu-1}-2)+1)}
\nonumber
\end {eqnarray}
The above expression is nonzero only if $\alpha_{\nu}=2,$ and
$p=m_1+m_2+\cdots +m_{\nu}.$ This means, that the summations over $m_i$ are not independent,
the condition
$$p=m_1+m_2+\cdots +m_{\nu}$$
confine the indices. Inserting the above result to Eq.
(\ref{ch330}) and taking into account the identity:
$$(4\nu-2p)!= 2^{(4\nu-2p)} (2\nu-p)!\pochh{1/2}{(2\nu-p)},$$
we reads:
\begin {eqnarray}
& &\left\{\ms C^{\nu}(\Lambda)\right\}_{p,\nu}\ =\
\frac{\nu!}{\pochh{1/2}{(2\nu-p)}} \sum_{\alpha_1=2}^4\ \cdots\
\sum_{\alpha_{\nu-1}=2}^4\  c_{\alpha_1}\cdots c_{\alpha_{\nu-1}}c_2
\sum\limits_{m_1=0}^{\alpha_1}\
\sum\limits_{m_2=0}^{\alpha_2}
\cdots\sum\limits_{m_{\nu-1}=0}^{\alpha_{\nu-1}}
\binom{\alpha_1}{m_1}\binom{\alpha_2}{m_2}\cdots \binom{2}{m_{\nu}} \nonumber \\
& &\frac{\Gamma(2(\nu-1)-p_1+1)}{\Gamma(2(\nu-1)-p_1-(\alpha_1-2)+1)}\cdots
\frac{\Gamma(2(\nu-i)-p_i+1)}{\Gamma(2(\nu-i)-p_i-(\alpha_i-2)+1)}\cdots
\frac{\Gamma(2-p_{\nu-1}+1)}{\Gamma(2-p_{\nu-1}-(\alpha_{\nu-1}-2)+1)}\nonumber \\
& &\sum\limits_{j_1=1}^{\Lambda-\nu}\
\sum\limits_{j_2=j_1+1}^{\Lambda-\nu+1}\ \cdots\
\sum\limits_{j_{\nu}=j_{(\nu-1)}+1}^{\Lambda}\ d^{m_1}_{j_1}\cdots\
d^{m_{\nu-1}}_{j_{\nu-1}}\ d^{m_{\nu}}_{j_{\nu}}\ Q_{j_{\nu}}^4\ Q_{j_{\nu}-1}^4\  \cdots\
Q_{j_2}^4\ Q_{j_1}^4\ .\nonumber
\end {eqnarray}
In above equation we used the definition:
$$p_i=p-m_1-\cdots-m_i.$$

We can simplify this formula by interchanging the order if the summations:
$$\sum_{\alpha_i=2}^4\sum\limits_{m_i=0}^{\alpha_i}\rightarrow \sum\limits_{m_i=0}^4\ \sum_{\alpha_i=\max(2,m_i)}^4$$
We can now provide the summations:
$$\Sigma(m_i,i,p_i) = \sum_{\alpha_i=\max(2,m_i)}^4\ c_{\alpha_i}\binom{\alpha_i}{m_i}\ \frac{\Gamma(2(\nu-i)-p_i+1)}{\Gamma(2(\nu-i)-p_i-(\alpha_i-2)+1)}.$$
The results are summarized in the following table:

\begin{table}[h]
\begin{center}
    \begin{tabular}{|c|c|}\hline
    $m_{i}$ & $\Sigma(m_i,i,p_i)$\\\hline
    0 & $(2(\nu-i)-p_{i}+1/2)(2(\nu-i)-p_{i}+3/2)$\\\hline
    1 & $4(2(\nu-i)-p_{i}+1/2)(2(\nu-i)-p_{i}+3/4)$\\\hline
    2 & $6(2(\nu-i)-p_{i})(2(\nu-i)-p_{i}-1)+9(2(\nu-i)-p_{i})+3/4$\\\hline
    3 & $4(2(\nu-i)-p_{i}-1/4)(2(\nu-i)-p_{i})$\\\hline
    4 & $(2(\nu-i)-p_{i}-1)(2(\nu-i)-p_{i})$\\\hline
    \end{tabular}
    \caption{Values of $\Sigma(m_j,j,p_j)$ for  $m_{i}$}
    \label{Hodnoty vseobecnej sumacie}
    \end{center}
    \end{table}

It is clear from this table that
$$\Sigma(m_{\nu},\nu,0) = c_2 \binom{2}{m_{\nu}}$$
    With this notation we finally can read for value $\left\{\ms C^{\nu}(\Lambda)\right\}_{p,\nu}:$
 \begin {eqnarray}
& &\left\{\ms C^{\nu}(\Lambda)\right\}_{p,\nu}\ =\
\frac{\nu!}{\pochh{1/2}{(2\nu-p)}}\ \sum\limits_{\{m_1,\ldots,m_\nu\}\atop m_1+\ldots+m_\nu=p}
\prod_{i=1}^{\nu}\Sigma(m_i,i,p_i)\\
& &\sum\limits_{j_1=1}^{\Lambda-\nu}\
\sum\limits_{j_2=j_1+1}^{\Lambda-\nu+1}\ \cdots\
\sum\limits_{j_{\nu}=j_{(\nu-1)}+1}^{\Lambda}\ d^{m_1}_{j_1}\cdots\
d^{m_{\nu-1}}_{j_{\nu-1}}\ d^{m_{\nu}}_{j_{\nu}}\ Q_{j_{\nu}}^4\ Q_{j_{\nu}-1}^4\  \cdots\
Q_{j_2}^4\ Q_{j_1}^4\ .\nonumber
\end {eqnarray}
For the continuum limit of the Eq. (\ref{ch309}) we reads:
\begin {eqnarray}
& &\lim_{\triangle \rightarrow 0}\left\{\triangle^{3\nu}\sum \limits_{p=0}^{2\nu}\ \frac{(\xi)^{2\nu-p}}{\left(\sigma\
Q_{N-1}Q_{N-2}\right)^{2\nu-p}}\ \left\{\ms
C^{\nu}(N-1)\right\}_{p,\nu}\right\}=\\
&=& \sum \limits_{p=0}^{2\nu}\ \left(\frac{c}{ Q^2(\beta)}\varphi_N^2\right)^{2\nu-p}\frac{\nu!}{\pochh{1/2}{(2\nu-p)}}\ \sum\limits_{\{m_1,\ldots,m_\nu\}\atop m_1+\ldots+m_\nu=p}
\prod_{i=1}^{\nu}\Sigma(m_i,i,p_i)\nonumber\\
&\times&\int_0^{\beta}d\tau_1\ \int_{\tau_1}^{\beta}d\tau_2\cdots
\int_{\tau_{\nu-1}}^{\beta}d\tau_{\nu}\ \
d^{m_1}(\tau_1)d^{m_2}(\tau_2)\cdots\ d^{m_{\nu}}(\tau_{\nu})\ Q^4(\tau_{\nu})\  \cdots\ Q^4(\tau_2)\
Q^4(\tau_1)\ .\nonumber
\end {eqnarray}
where:
$$d(\tau) = \frac{1}{2\gamma}\left(\coth(\gamma \tau)-\coth(\gamma \beta)\right),$$
$$Q(\tau) = 2\sinh(\gamma \tau).$$
and $\Sigma(m_i,i,p_i)$ are given in Table \ref{Hodnoty vseobecnej sumacie}.

\section{ Algebra of integrals.}

 For product of integrals with three indexes it can be red:
    \begin {equation}
    I_{\alpha,\beta,\gamma}(\tau) I\underbrace{_{a,\ldots,a}}_{n-3}(\tau)=\sum_{j=1}^{n-2}\sum_{k=j+1}^{n-1}\sum_{l=k+1}^{n} I\underbrace{_{a,\ldots,a,\alpha_j,a,\ldots,a,\beta_k,a,\ldots,a,\gamma_l,a,\ldots,a}}_{n}(\tau)
    \end {equation}
    For evaluation of the an-harmonic correction (\ref{ch342}) we need the following terms characterized by different algebraic
    factors in the sum on the right hand side:
    \begin {equation}
    I_{\alpha,\beta,a}(\tau) I\underbrace{_{a,\ldots,a}}_{n-3}(\tau)=\sum_{j=1}^{n-1}\sum_{k=j+1}^{n}(n-k) I\underbrace{_{a,\ldots,a,\alpha_j,a,\ldots,a,\beta_k,a,\ldots,a}}_{n}(\tau)
    \end {equation}
    The expansion of the summations over summation indices $j$ up to $n-1$ and $k$ up to $n$ was done by adding the zero terms due the
    factor $(n-k).$

    For two indexes $a$ we find:
    \begin {equation}
    I_{\alpha,a,a}(\tau) I\underbrace{_{a,\ldots,a}}_{n-3}(\tau)=\sum_{i=1}^{n-2}\sum_{j=i+1}^{n-1}\sum_{k=j+1}^{n}
    I\underbrace{_{a,\ldots,a,\alpha_{i},a,\ldots,a}}_{n}(\tau)
    \end {equation}
    Because the integrals $I\underbrace{_{a,\ldots,a,\alpha_{i},a,\ldots,a}}_{n}(\tau)$ are independent of summations indexes
     $j,k$ we find:
     \begin {equation}
    I_{\alpha,a,a}(\tau) I\underbrace{_{a,\ldots,a}}_{n-3}(\tau)=
    \sum_{i=1}^{n}\binom{n-i}{2}I\underbrace{_{a,\ldots,a,\alpha_{i},a,\ldots,a}}_{n}(\tau).
    \end {equation}
    We expand the summation for $i=n-1, i=n,$ because the added terms are zero.

    When index $a$ is in between $\alpha,$ and $\beta,$ we have:
    \begin {equation}
    I_{\alpha,a,\beta}(\tau) I\underbrace{_{a,\ldots,a}}_{n-3}(\tau)=\sum_{j=1}^{n-2}\sum_{k=j+1}^{n-1}\sum_{l=k+1}^{n} I\underbrace{_{a,\ldots,a,\alpha_j,a,\ldots,a,\beta_l,a,\ldots,a}}_{n}(\tau)
    \end {equation}
    Due to $k$ independence of the integral, we can red:
     \begin {equation}
    I_{\alpha,a,\beta}(\tau) I\underbrace{_{a,\ldots,a}}_{n-3}(\tau)=\sum_{j=1}^{n-2}\sum_{l=j+2}^{n} (l-1-j) I\underbrace{_{a,\ldots,a,\alpha_j,a,\ldots,a,\beta_l,a,\ldots,a}}_{n}(\tau)
    \end {equation}
    Thanks to factor $(l-1-j)$ we can expand the summations up to $j=n-1$ and $l=j+1:$
    \begin {equation}
    I_{\alpha,a,\beta}(\tau) I\underbrace{_{a,\ldots,a}}_{n-3}(\tau)=\sum_{j=1}^{n-1}\sum_{l=j+1}^{n} (l-1-j) I\underbrace{_{a,\ldots,a,\alpha_j,a,\ldots,a,\beta_l,a,\ldots,a}}_{n}(\tau)
    \end {equation}
For product of the integrals with four greek indexes we have:
    \begin {equation}
    I_{\alpha,\beta,\gamma,\delta}(\tau)I\underbrace{_{a,\ldots,a}}_{n-4}(\tau)= \sum_{i=1}^{n-3}\sum_{j=i+1}^{n-2}\sum_{k=j+1}^{n-1}\sum_{l=k+1}^{n}
    I\underbrace{_{a,\ldots,a,\alpha_{i},a,\ldots,a,\beta_{j},a,\ldots,a,\gamma_k,a,\ldots,a,\delta_l,a,\ldots,a}}_{n}(\tau)
    \end {equation}
    For the current evaluations there are interesting the  two special cases, when two greek indices are equal to $a$:
    \begin {equation}
    I_{\alpha,a,\gamma,a}(\tau)I\underbrace{_{a,\ldots,a}}_{n-4}(\tau)= \sum_{i=1}^{n-3}\sum_{j=i+1}^{n-2}\sum_{k=j+1}^{n-1}\sum_{l=k+1}^{n}
    I\underbrace{_{a,\ldots,a,\alpha_{i},a,\ldots,a,\gamma_k,a,\ldots,a}}_{n}(\tau).
    \end {equation}
   Because the integrals $I\underbrace{_{a,\ldots,a,\alpha_{i},a,\ldots,a,\gamma_k,a,\ldots,a}}_{n}(\tau)$ are independent of the summation indices
    $j,l,$ we have:
    \begin {equation}
    I_{\alpha,a,\gamma,a}(\tau)I\underbrace{_{a,\ldots,a}}_{n-4}(\tau)= \sum_{i=1}^{n-3}\sum_{k=i+2}^{n-1}(n-k)(k-i-1)I\underbrace{_{a,\ldots,a,\alpha_{i},a,\ldots,a,\gamma_k,a,\ldots,a}}_{n}(\tau)
    \end {equation}
    Due to the factor $(n-k)(k-i-1)$ we can expand the sum for index $k=i+1$ and $k=n$ also. Thanks to identity:
    $$k-i-1 = (n-i-2)-(n-k-1)$$
    and the new algebraic factors elouding us to extend the sum over the index $i$ we have:
    \begin{eqnarray*}
    I_{\alpha,a,\gamma,a}(\tau)I\underbrace{_{a,\ldots,a}}_{n-4}(\tau)&=& \sum_{i=1}^{n-1}\sum_{k=i+1}^{n}(n-k)(n-i-2)I\underbrace{_{a,\ldots,a,\alpha_{i},a,\ldots,a,\gamma_k,a,\ldots,a}}_{n}(\tau)-\\
    &-&\sum_{i=1}^{n-1}\sum_{k=i+1}^{n}(n-k)(n-k-1)I\underbrace{_{a,\ldots,a,\alpha_{i},a,\ldots,a,\gamma_k,a,\ldots,a}}_{n}(\tau)
    \end{eqnarray*}
    As the last example we evaluate the product:
    \begin {equation}
    I_{\alpha,\beta,a,a}(\tau)I\underbrace{_{a,\ldots,a}}_{n-4}(\tau)= \sum_{i=1}^{n-3}\sum_{j=i+1}^{n-2}\sum_{k=j+1}^{n-1}\sum_{l=k+1}^{n}
    I\underbrace{_{a,\ldots,a,\alpha_{i},a,\ldots,a,\beta_{j},a,\ldots,a}}_{n}(\tau)
    \end {equation}
    The integrals $I\underbrace{_{a,\ldots,a,\alpha_{i},a,\ldots,a,\beta_{j},a,\ldots,a}}_{n}(\tau)$ are independent of summation indexes
    $k$ and $l,$ which give to rise the factor $\binom{n-j}{2}.$ This factor aloud us to extend the summation over indexes $i$ and $j,$
    so in that way we have:
    \begin {equation}
    I_{\alpha,\beta,a,a}(\tau)I\underbrace{_{a,\ldots,a}}_{n-4}(\tau)= \sum_{i=1}^{n-1}\sum_{j=i+1}^{n}\binom{n-j}{2}
    I\underbrace{_{a,\ldots,a,\alpha_{i},a,\ldots,a,\beta_{j},a,\ldots,a}}_{n}(\tau)
    \end {equation}

For purposes of this article we don't need to evaluate the another identities.

\end{document}